\documentclass[paper,prd,reprint,preprintnumbers,nofootinbib,notitlepage,showpacs,showkeys]{revtex4-1}
\usepackage{latexsym,graphicx,amssymb,amsmath,mathrsfs} 
\usepackage{slashed}

\DeclareSymbolFont{bbold}{U}{bbold}{m}{n}
\DeclareSymbolFontAlphabet{\mathbbold}{bbold}

\newcommand{\be}{\begin{equation}}      
\newcommand{\ee}{\end{equation}}      
\newcommand{\bea}{\begin{eqnarray}}      
\newcommand{\eea}{\end{eqnarray}}    
     
\newcommand{\rt}[1]{{}}

\newcommand{\Tr}{\,\textrm{Tr}\,}

\newcommand{\ife}{\,\textrm{if}\,}
\newcommand{\nucl}{\,\textrm{nucl}\,}
\newcommand{\orr}{\,\textrm{or}\,} 
 
\newcommand{\ns}{\,\textrm{ns}\,}
\newcommand{\s}{\,\textrm{s}\,} 
\renewcommand{\min}{\,\textrm{min}\,} 
\newcommand{\GeV}{\,\textrm{GeV}\,} 
\newcommand{\MeV}{\,\textrm{MeV}\,} 
\newcommand{\els}{\,\textrm{else}\,} 
\newcommand{\eff}{\,\textrm{eff}\,} 
\newcommand{\ch}{\,\textrm{ch}\,} 
\newcommand{\liq}{\,\textrm{liq}\,} 
\newcommand{\fm}{\,\textrm{fm}\,} 
\newcommand{\cep}{\,\textrm{cep}\,} 
\newcommand{\const}{\,\textrm{const.}\,} 
\newcommand{\higher}{\,\textrm{higher order terms}\,}

\makeatletter
\renewcommand\appendix{\par
\setcounter{section}{0}%   
\setcounter{subsection}{0}% 
\gdef\thesection{\appendixname\space\@Alph\c@section}}

\long\def\unmarkedfootnote#1{{\long\def\@makefntext##1{##1}\footnotetext{#1}}}
\makeatother

\begin{document} 

\title{Axial anomaly and hadronic properties in a nuclear medium}
\author{G. Fej\H{o}s$^1$}
\email{fejos@rcnp.osaka-u.ac.jp}
\author{A. Hosaka$^{1,2}$}
\email{hosaka@rcnp.osaka-u.ac.jp}
\affiliation{$^1$Research Center for Nuclear Physics, Osaka University, Ibaraki, Osaka 567-0047, Japan}
\affiliation{$^2$Advanced Science Research Center, Japan Atomic Energy Agency, Tokai, Ibaraki, 319-1195 Japan}

\begin{abstract}
{We investigate meson and nucleon dynamics at finite baryon density and temperature by coupling the nucleon field and the omega meson to the three-flavor linear sigma model and calculate hadronic properties around the nuclear liquid-gas transition. We apply the functional renormalization group method, and find that mesonic fluctuations increase the strength of the coefficient of the $U_A(1)$ breaking determinant operator as a function of the chiral condensate. As a consequence, we find that the actual value of the anomaly increases discontinuously at the first order nuclear liquid-gas transition. We calculate how mesonic masses and partial restoration of chiral symmetry are modified due to such an effect.}
\end{abstract}

\maketitle

\section{Introduction}

Understanding strongly interacting dynamics in dense nuclear matter is a great challenge in nuclear many-body theory. At present times, controlled results on thermodynamics of the fundamental theory of quantum chromodynamics (QCD) are unlikely to be acquired for baryochemical potentials $\mu_B \gtrsim T$ \cite{borsanyi15}, and therefore, one is left without first principle calculations regarding the behavior of cold nuclear matter at finite density. Related phenomena, however, continuously receives huge attention. The equation of state (EOS) of cold dense nuclear matter found the deep interior of neutron stars (NSs) has been under several constrains since the discovery of two solar mass objects \cite{demorest10,antoniadis13}, and the recent measurement of gravitational waves of a NS-NS merger \cite{abbott17}. The hadron spectrum, which is expected to be modified due to partial restoration of chiral symmetry in the nuclear medium has also been of considerable interest, in particular regarding meson-nucleon bound states, which is regarded as a unique possibility of probing in-medium meson properties \cite{tanaka18}. The $\eta'$-nucleon interaction has in particular received attention, in that if the mass of the $\eta'$ particle dropped about $100 \MeV$ at normal nuclear density \cite{costa05,nagahiro06,sakai13,sakai16}, similarly to the $\Lambda(1405)$ $\bar{K}$-nucleon bound state, one might have the chance of observing an $\eta'$-nucleon composite. Spectroscopy experiments of the $^{12}C(\gamma,p)$ reaction via photon beam have been proposed by the LEPS2 collaboration (SPring-8 facility) \cite{nuramatsu13} and by the BGO-OD (ELSA accelerator). Furthermore, at the JAEA Heavy Ion (HI) project it is aimed to create conditions similar to that of a neutron star, i.e. $5$$-$$10$ times normal nuclear density, for the first time in a laboratory setting \cite{j-parc-hi16}. The program will expectedly help better understand and determine the QCD phase structure, e.g. critical points, phase boundaries and the EOS of nuclear matter.

Because of lack of first principle calculations, one applies effective models, which are based on (approximate) chiral symmetry of QCD. The most popular ones are the two- or three-flavor linear sigma models, the Nambu-Jona-Lasinio model, extended with the Polyakov loop, vector mesons and the nucleon \cite{parganlija13,herbst14,mitter14,kovacs16,rennecke16,almasi17,aoki18, floerchinger12,drews16}. Typically, effective models are not weakly coupled, therefore, one needs to go beyond mean field and perturbative approximations, and consider fluctuations to be important. Promising approaches for taking them into account are e.g. various functional methods, such as the Dyson-Schwinger technique \cite{eichmann16}, the 2PI method \cite{marko13,marko15} or the functional renormalization group (FRG) \cite{berges02,kopietz}. In this paper we extend our earlier formulations \cite{fejos16,fejos17} and apply the FRG scheme. Through this method, it is possible to extract nonperturbative information out of the corresponding models by deriving a scale evolution equation for the quantum effective action. An important aspect of the formulation is that it opens up the possibility of calculating field dependent coupling constants (functions), which can be interpreted as partial resummation of combinations of operators reflecting chiral symmetry.

This is in particular interesting for the $U_A(1)$ breaking `t Hooft determinant term, which describes the axial $U_A(1)$ anomaly. A closely related question regarding the QCD phase diagram and the physics of the $\eta'$ meson is the fate of this anomaly at finite temperature and density. In \cite{fejos17} we calculated how the anomaly coefficient changes in the three-flavor linear sigma model extended with nucleons, as quantum, thermal and density fluctuations are integrated out. We obtained that on the one hand, as hinted above, the anomaly acquired condensate dependence, and on the other hand it became temperature and density dependent. Due to the combination of these effects we found that as one raises the baryochemical potential and/or temperature, and thus as the chiral condensate starts to melt, the anomaly shows (a possibly intermediate) strengthening, before reaching a regime, where perturbative instanton calculations of QCD are applicable \cite{schaefer98}, and where the anomaly should gradually disappear. 

Our findings showed that one should be more cautious regarding a possible mass drop of the $\eta'$ particle, which was predicted by mean field calculations in the Nambu-Jona-Lasinio \cite{costa05,nagahiro06} and linear sigma models \cite{sakai13,sakai16}. A shortcoming of our earlier study, however, was that it could not distinguish between the nuclear liquid and gas phases, and it was not able to display the corresponding first order transition. In this paper we are looking at the system more carefully, and investigate to what extent the nuclear transition affects the anomaly, the mesonic spectrum and the partial restoration of chiral symmetry. In order to do so, we include a neutral vector meson, the $\omega$ particle into the system, which is modeling the repulsive short range interaction between the nucleons, and which is indispensable for proper description of the liquid-gas transition. For asymmetric nuclear matter, the $\rho$ meson is also necessary, but in this study we restrict ourselves to isospin symmetry.

The paper is organized as follows. In Sec. II, we introduce the model and the FRG formulation. We go into the details of how our approximation scheme is built up, and how model parameters are obtained from appropriate inputs. In Sec. III, we discuss the results and show various plots. Section IV contains the summary and outlook.

\section{Chiral nucleon-meson model}

\subsection{Basics}

The system we are interested in consists of the pseudoscalar and scalar meson ($M$) nonets in a nuclear [$\psi^T_N=(p, n)$], isospin symmetric environment. The nucleon-nucleon repulsive interaction is modeled by a vector particle ($\omega$), and we are to describe how the system behaves at finite temperature and baryochemical potential. The Euclidean Lagrangian takes the following form:
\bea
\label{Eq:Lag}
{\cal L}&=&\Tr [\partial_i M^\dagger \partial_i M]+V_{ch}[M]+a(\det M^\dagger + \det M) \nonumber\\
&-&\Tr [H(M^\dagger + M)] + \frac14 \omega_{ij}\omega_{ij}+\frac12 m_\omega^2 \omega_i\omega_i \nonumber\\
&+&\bar{\psi}_N(\partial\!\!\!/-\mu_B \gamma_0+g_Y \tilde{M}_5-ig_\omega \omega\!\!\!/)\psi_N,
\eea
where the potential term $V_{ch}[M]$ reflects chiral symmetry:
\bea
\label{Eq:Vch}
\!\!V_{ch}[M]&=&m^2 \Tr[M^\dagger M]+\lambda_1 \big(\Tr[M^\dagger M]\big)^2 \nonumber\\
&+&\lambda_2 \Tr[M^\dagger MM^\dagger M]+\higher.
\eea
(We will come back to the role of the higher order terms.) The mesonic field can be written as
\bea
M=\sum_{a=0}^8 (s^a+i\pi^a)T^a,
\eea
where $T^a=\hat{\lambda}^a/2$ are $U(3)$ generators ($\hat{\lambda}^a$ being the Gell-Mann matrices), while $s^a$ and $\pi^a$ correspond to the scalar and pseudoscalar mesons. The term in (\ref{Eq:Lag}) containing determinants describes the $U_A(1)$ anomaly, while $H=h_0T_0+h_8T_8$ is responsible for an explicit symmetry breaking. In order to couple the nonstrange (ns) nucleon field $\psi_N$ into the three-flavor meson model, we need to introduce a $\tilde{M}$ field, which belongs to a $U(2)$ subgroup of $U(3)$, which is spanned by some generators $\tilde{T}^a$:
\bea
\tilde{M}=\sum_{a=\ns,1,2,3} (s^a+i\pi^a)\tilde{T}^a.
\eea
Note that the nonstrange generator refers to the combination $T^{\ns}=\sqrt{2/3}T^0+1/\sqrt3 T^8$, but the $\tilde{T}^a$ matrices should be considered as $2\times 2$ (the $T^a$ ones are $3\times 3$). For symmetry reasons, when coupling the nucleons to the mesons one uses
\bea
\tilde{M}_5=\sum_{a=\ns,1,2,3} (s^a+i\pi^a\gamma_5)\tilde{T}^a,
\eea
where $\gamma_5$ is the fifth Dirac matrix. Finally, $\omega_{ij}=\partial_i \omega_j - \partial_j \omega_i$. Since fluctuations of $\omega$ will not be considered, and it will only serve as a background field, we may rescale $\omega_i \rightarrow \omega_i/g_\omega$, and thus at this point the model parameters are $m^2$, $\lambda_1$, $\lambda_2$, $a$, $h_0$, $h_8$, $g_Y$, and $G_\omega:=g_\omega^2/m_\omega^2$ (we will have some more due to higher order terms in $V_{ch}$). From (\ref{Eq:Lag}), the classical potential is the following:
\bea
\label{Eq:Vcl}
V_{cl}(M,\omega,\psi_N)&=&V_{ch}(M)+a(\det M^\dagger + \det M) \nonumber\\
&-&\Tr [H(M^\dagger + M)]-\frac{\omega^2}{2G_{\omega}} \nonumber\\
&+&\bar{\psi}_N(g_Y \tilde{M}_5-(\mu_B+\omega)\gamma_0)\psi_N,
\eea
where we assumed that the (Euclidean) timelike component of the $\omega_i$ field acquires an expectation value as $\langle \omega_i \rangle=-i\omega\delta_{i4}$. In what follows we take into account fluctuation corrections of $M$ and $\psi_N$ to (\ref{Eq:Vcl}) and calculate the effective potential of the system using the functional RG method. 

In the FRG, the classical action corresponding to the Lagrangian (\ref{Eq:Lag}) serves as a starting point of the renormalization group flows, defined at some ultraviolet (UV) scale $\Lambda$. Having in mind that (\ref{Eq:Lag}) is an effective theory valid up to scales of ${\cal O}$(1 GeV), we choose $\Lambda = 1 \GeV$. Using the compact notation of the fields, $\Phi=(M,\psi_N,\omega_i)$ (and defining the corresponding sources as $J$), the scale-dependent quantum effective action $\Gamma_k$ is
\bea
\Gamma_k[\Phi]&=&-\log Z_k[J]-\int J\Phi-\int \Phi^\dagger {\cal R}_k \Phi, \nonumber\\
Z_k[J]&=&\int {\cal D}\Phi e^{-(\int {\cal L}+\int J\cdot\Phi+\int \Phi^\dagger {\cal R}_k \Phi)}.
\eea
$\Gamma_k$ obeys the following flow equation:
\bea
\label{Eq:flow}
\partial_k \Gamma_k = \frac{1}{2} \int_p \int_q \Tr[(-2)^F(\Gamma_k^{(2)}+{\cal R}_k)^{-1}(q,p) \partial_k {\cal R}_k(p,q)], \nonumber\\
\eea
where $F=1$ for indices of the trace that belongs to fermionic variables, and $F=0$ for bosons. $\Gamma_k^{(2)}$ is the second derivative matrix of $\Gamma_k$ with respect to the fields, and ${\cal R}_k$ is a regulator matrix. In this paper, we choose for bosonic eigenmodes in ${\cal R}_k$ the entries
\bea
\label{Eq:RegB}
R^B_k(q,p)&=&R^B_k({\bf q})\delta(q+p)\nonumber\\
&=&(k^2-{\bf q}^2)\Theta(k^2-{\bf q}^2)\delta({q}+{p}),
\eea
and
\bea
\label{Eq:RegF}
R^F_k(q,p)&=&R^F_k({\bf q})\delta(q+p)\nonumber\\
&=&i{\bf q}\!\!\!/\Big(\sqrt{\frac{k^2}{{\bf q^2}}}-1\Big)\Theta(k^2-{\bf q}^2)\delta({q}+{p})
\eea
for fermionic ones, where boldfaced variables are $3$-momenta. It is easy to show that, as mentioned above, $\Gamma_{k=\Lambda} = \int {\cal L}$, and that $\Gamma_{k=0}=\Gamma_{1PI}$, the latter being the ordinary 1PI quantum effective action.

\subsection{Approximation scheme}

The flow equation (\ref{Eq:flow}) can only be solved in approximation schemes. 
From now on, we neglect all scale dependence of the kinetic terms (i.e. wave function renormalizations) in $\Gamma_k$, and since we are only interested in homogeneous field configurations, we rather use the effective potential $V_{\eff\!,k}$, defined through $\Gamma_k|_{\Phi = \const} = \int_x V_{\eff\!,k}$. The ansatz for $V_{\eff\!,k}$, based on (\ref{Eq:Vcl}) is as follows \cite{fejos16}:
\bea
\label{Eq:Vkans}
V_{\eff\!,k}[M,\omega,\psi_N]&=&V_{\ch\!,k}(M) \nonumber\\
&+&A_k(M)\cdot (\det M^\dagger + \det M)\nonumber\\
&-&\Tr [H(M^\dagger + M)]-\frac{\omega^2}{2G_\omega} \nonumber\\
&+&\bar{\psi}_N(g_Y \tilde{M}_5-(\mu_B+\omega)\gamma_0)\psi_N,
\eea
where we are only interested how the chiral potentials evolve via the RG flow, and neglected the flows of the Yukawa couplings $g_Y$ and $G_\omega$, and also that of the explicit symmetry breaking term proportional to $H\equiv h_0T^0+h_8T^8$. These simplifications are expected to be good approximations at low enough temperatures \cite{rennecke16}. The ansatz (\ref{Eq:Vkans}) has to be compatible with the flow equation (\ref{Eq:flow}), therefore, we have to split $V_{\ch\!,k}$ into two parts:
\bea
V_{\ch\!,k}(M)=V_{k}(M)+\tilde{V}_{k}(\tilde{M}),
\eea
where $V_k$ reflects $U(3)\times U(3)$ symmetry, while $\tilde{V}_{k}$ is invariant under $U(2)\times U(2)$ two-flavor rotations. The necessity of such a splitting is due to the fact that nucleon fluctuations do not contribute to the strange sector, thus their presence has to introduce a part of $V_{\eff\!,k}$ that only reflects a two-flavor chiral symmetry, which is generated by the nonstrange and isospin matrices.

Using (\ref{Eq:Vkans}), the flow equation (\ref{Eq:flow}) turns into the following form for $V_{\eff\!,k}$ at finite temperature $T$:
\bea
\label{Eq:flowVk}
\partial_k&&\!\!\!\!\!\!\! V_{\eff\!,k}=\nonumber\\
&&\frac12 \tilde{\partial}_k T\sum_n\int_{\bf q} \Tr \log \big[\Omega_n^2+{\bf q}^2+V_{\eff\!,k}^{B(2)}+R_k^B({\bf q})]\nonumber\\
&-&\tilde{\partial}_k T\sum_{n}\int_{\bf q} \Tr \log \big[\tilde{\Omega}_n^2+{\bf q}^2+V_{\eff\!,k}^{F(2)}+R_k^F({\bf q})], \nonumber\\
\eea
where $\Omega_n=2n \pi T$, $\tilde{\Omega}_n=(2n+1)\pi T$ are bosonic and fermionic Matsubara frequencies, respectively, and $\tilde{\partial}_k$ acts only on the regulators. We separated the second derivative matrix $V_{\eff\!,k}^{(2)}$ into bosonic $(V_{\eff\!,k}^{B(2)})$ and fermionic $(V_{\eff\!,k}^{F(2)})$ parts [one may think of them as matrices satisfying $V_{\eff\!,k}^{(2)}=(V_{\eff\!,k}^{B(2)})\oplus(V_{\eff\!,k}^{F(2)})$]. Note that, as mentioned already, fluctuations of $\omega$ are not taken into account, therefore in $V_{\eff\!,k}^{B(2)}$ only second derivatives of the fields of $M$ are present.

Note that
\bea
\partial_k V_{\eff\!,k} &=& \partial_k V_k (M) + \partial_k \tilde{V}_k (\tilde{M}) \nonumber\\
&+& \partial_k A_k(M)\cdot (\det M^\dagger + \det M).
\eea
We need projections of the rhs of (\ref{Eq:flowVk}) to get individual equations for the scale evolution of $V_k$, $\tilde{V}_k$ and $A_k$. Concerning $\tilde{V}_{k}$ first, obviously the second term of (\ref{Eq:flowVk}) gives the leading order, but note that it backreacts on $V_{\eff\!,k}^{B(2)}$, and provides subleading contributions too. We do not take these into account and, therefore,
\bea
\partial_k \tilde{V}_k=-&T \sum_{n}\tilde{\partial}_k \int_{\bf q} \Tr \log \big[\tilde{\Omega}_n^2+{\bf q}^2+V_{\eff\!,k}^{F(2)}+R_k^F({\bf q})]. \nonumber\\
\eea
Performing the momentum integral and dropping terms that would contribute to the flow of the explicit breaking, we arrive at
\bea
\label{Eq:Vktildeflow}
\partial_k \tilde{V}_k=-\frac{2k^4T}{3\pi^2}\sum_{n} \Tr \frac{1}{(\tilde{\Omega}_n-i\mu_B^{(\eff)})^2+k^2+g_Y^2 \tilde{M}^\dagger \tilde{M}},\nonumber\\
\eea
where $\mu_B^{(\eff)}$ is an effective baryochemical potential corrected by $\omega$: $\mu_B^{(\eff)}=\mu_B+\omega$. A possible way to calculate the trace in (\ref{Eq:Vktildeflow}) is to substitute
\bea
g_Y^2 \tilde{M}^\dagger \tilde{M}\quad \rightarrow \quad g_Y^2\Delta \hat{I} + \frac{g_Y^2}{2}\Tr(\tilde{M}^\dagger \tilde{M}),
\eea
and perform an expansion in terms of $\Delta \hat{I}\equiv g_Y^2(\tilde{M}^\dagger \tilde{M}-\frac12\Tr(\tilde{M}^\dagger \tilde{M}))$:
\bea
\partial_k \tilde{V}_k(\tilde{M})=-\frac{2k^4}{3\pi^2}&&T\sum_{n}\Bigg[\frac{2}{\tilde{\omega}_n^2+E_k^2} \nonumber\\
&&+\sum_{m=1}^\infty (-1)^m \Big(\frac{g_Y}{2}\Big)^m \frac{\Tr (\Delta \hat{I})^m}{(\tilde{\omega}_n^2+E_k^2)^m}\Bigg], \nonumber\\
 \eea 
where $\tilde{\omega}_n=\tilde{\Omega}_n-i\mu_B^{(\eff)}$, $E_k^2=k^2+m_N^2$, and $m_N^2=\frac{g_Y^2}{2}\Tr(\tilde{M}^\dagger \tilde{M})$ is corresponding to the nucleon mass. Note that for a background where $\tilde{M} \sim {\bf 1}$ (i.e. in the presence of a nonstrange diagonal condensate), $\Delta \hat{I}\equiv 0$, therefore, for our purposes it is enough to consider
\bea
\label{Eq:Vktildeflowfinal}
\partial_k \tilde{V}_k(\tilde{M}) &=& -\frac{2k^4}{3\pi^2}T\sum_{n} \frac{2}{\tilde{\omega}_n^2+E_k^2}\nonumber\\
&=&-\frac{k^4}{3\pi^2E_k}\sum_{\pm}\Big[\coth \Big(\frac{E_k\pm\mu_B^{(\eff)}}{2T}\Big)\Big].
\eea
Note that since the flow of $g_Y$ is neglected, one can integrate (\ref{Eq:Vktildeflowfinal}) and get
\bea
\label{Eq:Vtildek}
\tilde{V}_{k=\Lambda}(\tilde{M}) -&& \tilde{V}_{k=0}(\tilde{M})=-\frac{1}{3\pi^2}\int_0^\Lambda dk \frac{k^4}{E_k}\nonumber\\
&&+\frac{2}{3\pi^2}\int_{0}^{\infty}\frac{k^4}{E_k}\sum_{\pm} n_F(E_k\pm \mu_B^{(\eff)}),
\eea
where $n_F(x)=(\exp(x/T)+1)^{-1}$ is the Fermi-Dirac distribution. Since the first term on the rhs of (\ref{Eq:Vtildek}) is an environment independent function of $\tilde{M}$, one may combine it with $\tilde{V}_{k=\Lambda}$ to introduce $\tilde{V}_{L}$ and arrive at
\bea
\tilde{V}_{k=0}(\tilde{M})=\tilde{V}_L(\tilde{M}) - \frac43 \int \frac{d^3k}{(2\pi)^3} \frac{k^2}{E_k}\sum_{\pm} n_F(E_k\pm \mu_B^{(\eff)}). \nonumber\\
\eea
Here $\tilde{V}_L$ is a chiral invariant function and plays the role of the (corrected) initial value of the flow, i.e. has to be of a $U(2)\times U(2)$ form of the classical potential (\ref{Eq:Vch}). One has to adjust its parameters in order to reproduce appropriate physical quantities in the vacuum (see the next subsection). The second term is just the standard one-loop contribution. It shows that without considering the flow of the Yukawa coupling $g_Y$, one does not go beyond perturbation theory.

The real strength of the FRG emerges when we consider the flows of $V_k(M)$ and $A_k(M)$. Details of obtaining them are worked out in detail in \cite{fejos16}, here we just briefly review the procedure. By definition,
\bea
\label{Eq:VkBflow}
\partial_k V_k &&+ \partial_k A_k (\det M^\dagger + \det M)=\nonumber\\
&&\frac12 \tilde{\partial}_k T\sum_n\int_q \Tr \log \big[\Omega_n^2+{\bf q}^2+V_{\eff\!,k}^{B(2)}(q)+R_k^B(q)], \nonumber\\
\eea
where $V_{\eff\!,k}^{B(2)}$, as explained before, contains only those contributions that reflect $U(3)\times U(3)$ symmetry (derivatives of $\tilde{V}_k$ do not count here).
First, one separates the flow of $V_k$. To reach that, $A_k=0$ is taken, and $V_k$ is approximated via a chiral invariant expansion \cite{fejos15}:
\bea
\label{Eq:chex}
V_k(M)=U_k(I_1)+C_k(I_1)\cdot I_2 + \dots
\eea
where
\begin{subequations}
\bea
I_1&=&\Tr (M^\dagger M), \\
I_2&=&\Tr (M^\dagger M - \Tr(M^\dagger M)/3)^2.
\eea
\end{subequations}
Projecting the flow equation onto a subspace where $I_2=0$ leads to the flow of $U_k(I_1)$, while after projecting (\ref{Eq:VkBflow}) onto the subspace of ${\cal O}(I_2)$, one obtains the flow of $C_k(I_1)$. Finally, we consider $A_k$ and perform one more projection, now onto the subspace of ${\cal O}(I_{\det})$, where $I_{\det}=\det M^\dagger + \det M$. These flow equations can be found in Appendix A, together with formulas that are helpful to obtain the $V_{\eff\!,k}^{B(2)}$ derivatives. 

Now we need to choose initial conditions for $U_{k=\Lambda}$, $C_{k=\Lambda}$, $A_{k=\Lambda}$ and $\tilde{V}_{k=\Lambda}$. We restrict ourselves to renormalizable operators in the three-flavor sector, and based on (\ref{Eq:Vch}), we choose
\bea
U_{k=\Lambda}&=&m^2 I_1 + (\lambda_1+\lambda_2/3) I_1^2, \nonumber\\
C_{k=\Lambda}&=&\lambda_2, \quad A_{k=\Lambda}=a.
\eea
At this point it has to be emphasized that, for $\tilde{V}_{k=\Lambda}$, we also need higher order contributions. These terms are nonrenormalizable, but given the fact that we are dealing with an effective theory, and the cutoff $\Lambda = 1 \GeV$ is rather small, one must not rule out the presence of such interactions. While one can perform physically meaningful parametrizations without introducing these types of terms in $V_{k=\Lambda}$ if the properties of nuclear matter are not of importance, here it turns out that one does need (in the RG sense) irrelevant operators for the two-flavor piece, $\tilde{V}_{k=\Lambda}$ (or $\tilde{V}_L$). The reason is that via $\tilde{V}_{k}$ the model has to be capable of describing the liquid-gas transition of nuclear matter, therefore, one needs a double-well potential to obtain the corresponding first order transition. This can only be achieved by not neglecting nonrenormalizable interactions at the UV scale. Keeping this in mind, the complete effective potential at this point takes the form of
\bea
\label{Eq:Veff1}
V_{\eff\!,k=0}[M,\omega]&=&U_{k=0}(I_1)+C_{k=0}(I_1)\cdot I_2\nonumber\\
&+&A_{k=0}(I_1)\cdot I_{\det}-h_ss_s-h_{\ns}s_{\ns}\nonumber\\
&+&\tilde{V}_L(\tilde{M})-\frac{\omega^2}{2G_\omega}, \nonumber\\
&-&\frac43\int \frac{d^3k}{(2\pi)^3}\frac{k^2}{E_k}\sum_{\pm}n_F(E_k\pm \mu_B^{(\eff)}), \nonumber\\
\eea
where we also performed a basis change in the $0$$-$$8$ sector and introduced nonstrange (ns) and strange (s) variables:
\bea
\label{Eq:snsss}
\begin{pmatrix}
s_{\ns} \\ s_{\s}
\end{pmatrix}
= \frac{1}{\sqrt3} \begin{pmatrix}
\sqrt2 & 1 \\
1 & -\sqrt2
\end{pmatrix}
\begin{pmatrix}
s_0 \\ s_8
\end{pmatrix},
\eea
and similarly for $(h_0, h_8) \leftrightarrow (h_{\ns},h_{\s})$. One notices that it is unnecessary after all to specify $\tilde{V}_{k=\Lambda}$, as only $\tilde{V}_L$ appears in (\ref{Eq:Veff1}). The condition that determines the latter function is that one should get an effective two-flavor description of the liquid-gas transition at $T=0$ (similarly as in \cite{floerchinger12,drews16}), after minimizing the effective potential with respect to $s_{\s}$. (We denote the minimum point by $s_{\s\!,\min}$, which is a function of $s_{\ns}$.) Keeping in mind that nonrenormalizable interactions can be present and tuned in $\tilde{V}_L(\tilde{M})$, we combine the first four terms with $\tilde{V}_L(\tilde{M})$ and let it equal a chiral expression, expanded around the nonstrange minimum in the vacuum that is eight order in the fields:
\bea
\label{Eq:VLdef}
&&U^{T=0}_{k=0}(I_1|_{s^{(T=0)}_{\s\!,\!\min}})+C^{T=0}_{k=0}(I_1|_{s^{(T=0)}_{\s\!,\!\min}})\cdot I_2|_{s^{(T=0)}_{\s\!,\!\min}}\nonumber\\
&&+A^{T=0}_{k=0}\cdot I_{\det}|_{s^{(T=0)}_{\s\!,\!\min}}-h_{\s}s^{(T=0)}_{\s\!,\!\min}+\tilde{V}_L(\tilde{M})\nonumber\\
&&\equiv \sum_{n=1}^4 b_n \Big(\tilde{I}_1-\frac12(v^{(T=0)}_{\ns\!,\!\min})^2\Big)^n,
\eea
where the $\{b_n\}$ $(n=1\dots 4)$ coefficients represent four new model parameters. $\tilde{I}_1$ is the analog of $I_1$: $\tilde{I}_1=\Tr(\tilde{M}^\dagger \tilde{M})/2$, and $v^{T=0}_{\ns\!,\!\min}$ is the true minimum of $s_{\ns}$ in the vacuum (i.e. the pion decay constant, as we will see shortly). Note that $I_1|_{s_{\s}\!,\!\min}$, $I_2|_{s_{\s}\!,\!\min}$ and $I_{\det}|_{s_{\s}\!,\!\min}$ has to be interpreted as functions of $\tilde{I}_1$, as we wish to obtain an effective two-flavor chiral invariance. Operators such as $\tilde{I}_2 \equiv \Tr\big(\tilde{M}^\dagger \tilde{M}-\Tr(\tilde{M}^\dagger \tilde{M})/2\big)^2$ should not appear due to our choice of (\ref{Eq:Vktildeflowfinal}), i.e. we are interested in field configurations where $\tilde{M} \sim {\bf 1}$. In a background of $s_{\ns}$, $\tilde{I}_1=s_{\ns}^2/2$, and therefore, in the minimum $s_{\s\!,\!\min}$, we may associate invariants with each other through the following identifications:
\bea
I_1|_{s_{\s\!,\!\min}}\rightarrow \tilde{I}_1+\frac{s^2_{\s\!,\!\min}}{2},&& \!\!\!\!I_2|_{s_{\s\!,\!\min}}\rightarrow \frac16 (\tilde{I}_1-s_{\s\!,\!\min}^2)^2, \nonumber\\
I_{\det}|_{s_{\s\!,\!\min}}&\rightarrow &\tilde{I}_1\frac{s_{\s\!,\!\min}}{\sqrt2}.
\eea
Expressing $\tilde{V}_L$ from (\ref{Eq:VLdef}), and adding an irrelevant constant to (\ref{Eq:Veff1}), $h_{\ns}s_{\ns\!,\!\min}^{(T=0)}$, for convenience, we arrive at
\begin{widetext}
\bea
\label{Eq:Vefffinal}
V_{\eff\!,k=0}&&\!\!\!\!\!\!\!(M,\omega)=U_{k=0}(I_1)+C_{k=0}\cdot I_2 + A_{k=0}(I_1)\cdot I_{\det} \nonumber\\
&-& \Big[ U^{T=0}_{k=0}(\tilde{I}_1+ s^{2(T=0)}_{\s\!,\!\min}/2)+C^{T=0}_{k=0}(\tilde{I}_1+ s^{2(T=0)}_{\s\!,\!\min}/2)\cdot \frac16 (\tilde{I}_1-s_{\s\!,\!\min}^{2(T=0)})^2+A_{k=0}^{T=0}(\tilde{I}_1+ s^{2(T=0)}_{\s\!,\!\min}/2)\cdot \tilde{I}_1\frac{s^{(T=0)}_{\s\!,\!\min}}{\sqrt2}\Big] \nonumber\\
&-&h_{\s} (s_{\s}-s_{\s\!,\!\min}^{(T=0)})-h_{\ns} (s_{\ns}-v_{\ns\!,\!\min}^{(T=0)})+\sum_{n=1}^4 b_n \Big(\tilde{I}_1-\frac12(v^{(T=0)}_{\ns\!,\!\min})^2\Big)^n -\frac{\omega^2}{2G_\omega}\nonumber\\
&-&\frac43\int \frac{d^3k}{(2\pi)^3}\frac{k^2}{E_k}\sum_{\pm}n_F(E_k\pm \mu_B^{(\eff)}).
\eea
\end{widetext}
Via this construction, after minimizing (\ref{Eq:Vefffinal}) with respect to $s_{\s}$ at $T=0$, we manage to have the following form of the effectively two-flavor potential:
\bea
\label{Eq:VeffT0}
\!\!\!\!\!\!\!\!\!\!V_{\eff\!,k=0}^{T=0}[\tilde{M},\omega]&=&\sum_{n=1}^4 b_n \Big(\tilde{I}_1-\frac12(v^{(T=0)}_{\ns\!,\!\min})^2\Big)^n \nonumber\\
&-&h_{\ns} (s_{\ns}-v_{\ns\!,\!\min}^{(T=0)})-\frac{\omega^2}{2G_\omega} \nonumber\\
&-&\frac43\int \!\!\!\frac{d^3k}{(2\pi)^3}\frac{k^2}{E_k}\sum_{\pm}n^{T=0}_F(E_k\pm \mu_B^{(\eff)}).
\eea
At this point it is important to mention that at $k=0$ not all $\{b_n\}$ coefficients are compatible with the solution of the flow equation, as at $k=0$ the effective potential has to be convex. In particular, we intend to choose $\{b_n\}$ such that it leads to a double-well potential, which should never come out as a result given the flow equation is solved numerically. We do not feel that it is of serious problem, because it has been shown in earlier works \cite{berges97,fukushima10} that there exists a critical scale $k_c$ beyond which position of the minima of a double-well potential does not change, and convexity is built up by flattening of the hill between those minima. Therefore, we think of (\ref{Eq:VeffT0}) as a construction which only models the positions of the minima, but not the structure in between. As a consequence, for example, we will not be using (\ref{Eq:VeffT0}) to calculate surface tension of a liquid droplet.

Now we are in a position to determine the model parameters. Note that from the three-flavor sector we have $m^2$, $g_1$, $g_2$, $h_0$, $h_8$ (or $h_{\ns}$ and $h_{\s}$), $a$, and $b_1$, $b_2$, $b_3$, $b_4$ from the additional two-flavor piece. Furthermore, one needs to determine $g_Y$ and $G_{\omega}$ Yukawa couplings. That is 12 parameters in total, which are to be dealt with in the next subsection.

\subsection{Parametrization}

We start the parametrization by recalling that the partially conserved axialvector current (PCAC) relations give
\bea
m_{\pi}^2f_{\pi}=h_{\ns}, \quad m_K^2f_K=\frac{h_{\ns}}{2}-\frac{h_s}{\sqrt2},
\eea
where $m_{\pi}^2=\partial^2 V^{T=0}_{\eff\!,k=0}/\partial \pi_i^2$ $[i=1,2,3]$ and $m_K^2=\partial^2 V^{T=0}_{\eff\!,k=0}/\partial \pi_j^2$ $[j=4,5,6,7]$. Using physical pion ($m_{\pi}=140 \MeV$) and kaon ($m_K=494 \MeV$) masses and decay constants ($f_{\pi}=93 \MeV$, $f_K=113 \MeV$), one gets
\bea
\label{Eq:h}
h_{\ns}&=&m_{\pi}^2f_{\pi} \approx (122 \MeV)^3, \nonumber\\
h_{\s}&=&\frac{1}{\sqrt2}(2m_K^2f_K-m^2_{\pi}f_{\pi}) \approx (335 \MeV)^3,
\eea
or
\bea
h_0&=&\sqrt{\frac23}\big(m_{\pi}^2f_{\pi}/2+m_K^2f_K)\approx (285 \MeV)^3, \nonumber\\
h_8&=&\frac{2}{\sqrt3}\big(m_{\pi}^2f_{\pi}-m_K^2f_K) \approx -(310 \MeV)^3.
\eea
Ward identities of chiral symmetry lead to
\begin{subequations}
\bea
\frac{\partial V^{T=0}_{\eff\!,k=0}}{\partial \pi_{i=1,2,3}}&=&m_{\pi}^2s_{\ns}-h_{\ns}, \\
\frac{\partial V^{T=0}_{\eff\!,k=0}}{\partial \pi_{i=4,5,6,7}}&=&\frac{m_K^2-m_{\pi}^2}{\sqrt2}s_{\ns}+m_K^2s_{\s}-h_{\s},
\eea
\end{subequations}
i.e. that no matter what the remaining parameters are, if we use (\ref{Eq:h}), in the minimum of the effective potential
\bea
v_{\ns\!,\!\min}^{(T=0)}=f_{\pi}, \quad v_{\s\!,\!\min}^{(T=0)}=\sqrt2(f_K-f_{\pi}/2).
\eea
Note that $s_{\s\!,\!\min}$ that we used earlier equals $v_{\s\!,\!\min}$ only when the nonstrange condensate is set to $f_\pi$: $s^{(T=0)}_{\s\!,\!\min}\big|_{s_{\ns}=f_{\pi}}=v_{\s\!,\!\min}^{(T=0)}$.

Now we make use of some of the zero temperature properties of nuclear matter. First one notes that the nucleon mass in the current model entirely comes from the spontaneous breaking of chiral symmetry, $m_{N}(s_{\ns})=g_Ys_{\ns}/2$, and since $m_{N}(f_{\pi})\approx 939 \MeV$ in the vacuum (i.e. at $s_{\ns}=v_{\ns\!,\!\min}^{(T=0)}\equiv f_{\pi}$), we get $g_Y\approx 20.19$. Normal nuclear density, $n_N \approx 0.17 \fm^{-3} \approx (109.131 \MeV)^3$ determines the Fermi momentum of the nucleons, as at $T=0$ one has $n_N=4\int_p n_F(\sqrt{p^2+m_N^2}-\mu_B^{(\eff)})|_{T=0}\equiv \frac{2}{3\pi^2}p_F^3$, thus $p_F \approx 267.9 \MeV \approx 1.36 \fm^{-1}$. This immediately leads to the value of the nonstrange condensate in the nuclear liquid phase, since the Landau mass, which is defined as $M_L=\sqrt{p_F^2+m_N^2(v_{\ns\!,\!\liq})}$ (note that $v_{\ns\!,\!\liq} \neq f_{\pi}$ as it corresponds to the liquid phase) is known to be $M_L \approx 0.8m_N(f_{\pi}) \approx 751.2 \MeV$, and therefore $v_{\ns\!,\!\liq} \approx 69.52 \MeV$. The Landau mass also determines the value of the $\omega$ condensate in the liquid phase (at the critical point), as it is nothing but the critical effective chemical potential: $M_L = \mu_{B,c} + \omega_c$, where the real chemical potential equals $\mu_{B,c}=m_{N}(f_{\pi})-B \approx 922.7 \MeV$, $B \approx 16.3 \MeV$ being the binding energy per nucleon. This calculation yields $\omega_c \approx -171.5 \MeV$. For the sake of an example, we can calculate the compression modulus $K$ of nuclear matter:
\bea
K=9\frac{n_N}{\partial n_N/\partial \mu_B}\bigg|_{T=0}=3\frac{M_L^2-m^2_N(v_{\ns\!,\!\liq})}{M_L},
\eea
for which we get $K\approx 287 \MeV$, in decent agreement with the experimentally established value \cite{stone14}.

All in all, we need to adjust $G_\omega$ and $b_i$ $(i=1,2,3,4)$ in (\ref{Eq:VeffT0}) such that a first order transition occurs at $T=0$, $\mu_B=922.7 \MeV$ from $s_{\ns}=f_{\pi}$ to $s_{\ns}=v_{\ns\!,\!\liq}$, while $\omega$ acquires its critical value $\omega_c$.

First, one minimizes (\ref{Eq:VeffT0}) at $T=0$ with respect to $\omega$:
\bea
\label{Eq:omegaeos}
&&\frac{\partial V_{\eff\!,k=0}^{(T=0)}[\tilde{M},\omega]}{\partial \omega}=-\frac{\omega}{G_\omega}+\nonumber\\
&&4\sum_{\pm}\int \!\!\frac{d^3p}{(2\pi)^3} \frac{\pm 1}{\exp[(\sqrt{p^2+m_{N}^2}\pm (\mu_B+\omega))/T]+1}\Bigg|_{T=0}\!\!\!. \nonumber\\
\eea
Setting the left-hand side of (\ref{Eq:omegaeos}) to zero and evaluating the integral we get
\bea
0=-\frac{\omega}{G_\omega}&-&\frac{1}{6\pi^2}\Big[(\mu_B+\omega)^2-g_Y^2s_{\ns}^2/4\Big]^{3/2} \nonumber\\
&\times&\Theta \big((\mu_B+\omega)^2-g_Y^2s_{\ns}^2/4)\big).
\eea
One solves this equation for $\omega=\omega(s_{\ns})$ and the constraint that determines $G_\omega$ is $\omega(v_{\ns,\nucl})=\omega_c$ at $\mu_B=\mu_{B,c}$. We get $G_\omega^{-1} \approx 7573.17 \MeV^2$.

Now, if we plug $\omega=\omega(s_{\ns})$ into $(\ref{Eq:omegaeos})$ numerically, we obtain an effective potential for $\tilde{M}$ (or $s_{\ns}$) only. This potential has to have the following properties:
\begin{subequations}
\label{Eq:constr}
\bea
\label{Eq:constr1}
V^{(T=0)}_{\eff\!,k=0}[\tilde{M}=f_{\pi}T^{\ns}]&=&0 \\
\label{Eq:constr2}
\frac{\partial V^{(T=0)}_{\eff\!,k=0}[\tilde{M}]}{\partial s_{\ns}}\Bigg|_{\tilde{M}=f_{\pi}T^{\ns}}&=&0, \\
\label{Eq:constr3}
V^{(T=0)}_{\eff\!,k=0}[\tilde{M}=v_{\ns\!,\!\liq}T^{\ns}]&=&0, \\
\label{Eq:constr4}
\frac{\partial V^{(T=0)}_{\eff\!,k=0}[\tilde{M}]}{\partial s_{\ns}}\Bigg|_{\tilde{M}=v_{\ns\!,\!\liq}T^{\ns}}&=&0.
\eea
\end{subequations}
Note that by construction (\ref{Eq:constr1}) and (\ref{Eq:constr2}) are automatically satisfied by (\ref{Eq:VeffT0}). In addition to (\ref{Eq:constr3}) and (\ref{Eq:constr4}), we require the pion mass to be physical, i.e. $m_{\pi}=140 \MeV$, as it is entirely determined by the two-flavor piece of the effective potential. Furthermore, we tune the parameters such that the critical end point of the nuclear transition is at $T_{\cep}=18 \MeV$ \cite{elliott13}. These conditions determine the $\{b_i\}$ parameters:
\bea
b_1/f_{\pi}^2 &\approx& 2.266, \quad b_2\approx 25.043, \nonumber\\
b_3\cdot f_{\pi}^2 &\approx& -12.572, \quad b_4 \cdot f_{\pi}^4\approx 169.312.
\eea

\begin{figure*}
\begin{center}
\raisebox{0.05cm}{
\includegraphics[bb = 0 490 520 770,scale=0.36,angle=270]{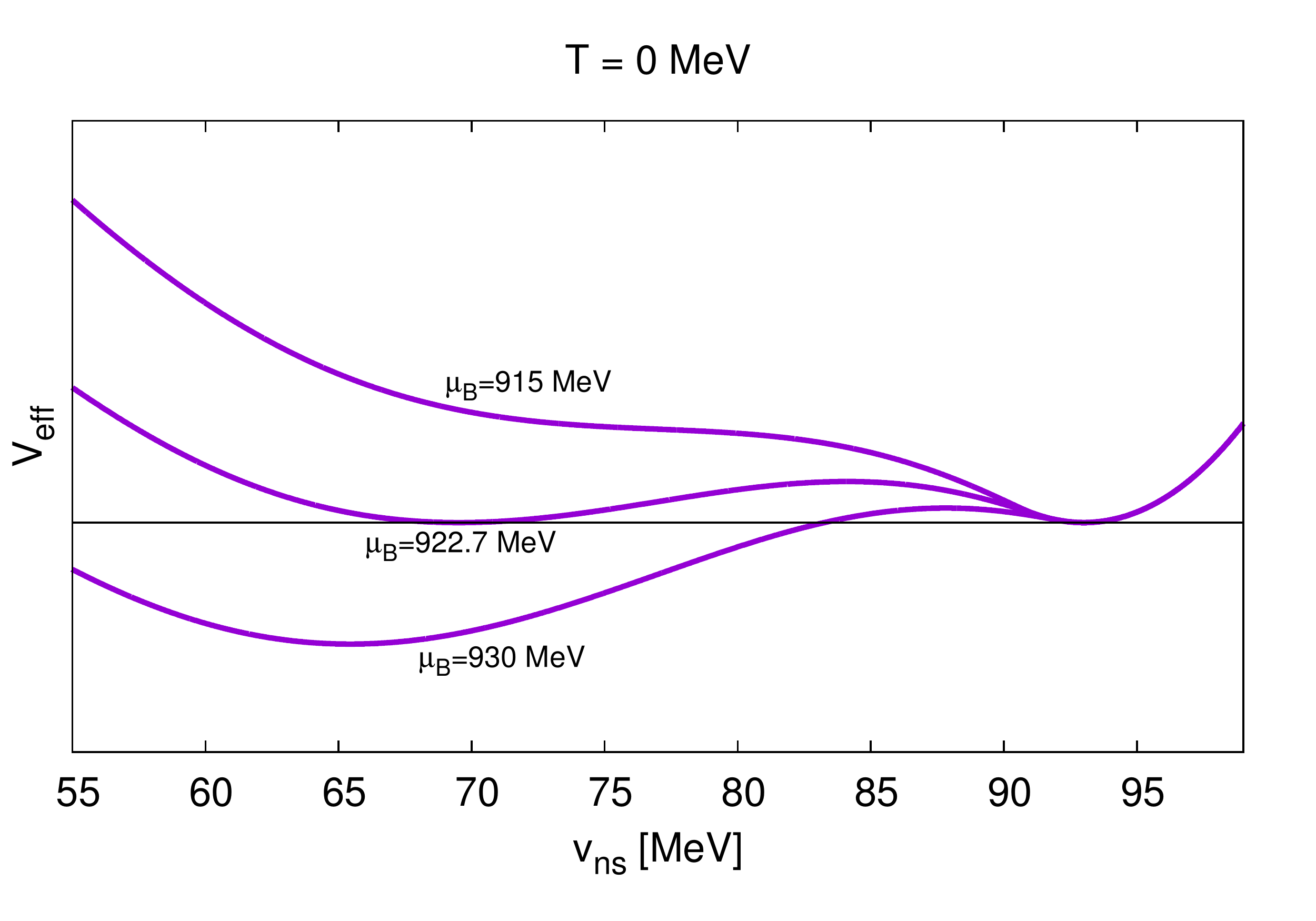}}
\includegraphics[bb = 0 80 520 170,scale=0.36,angle=270]{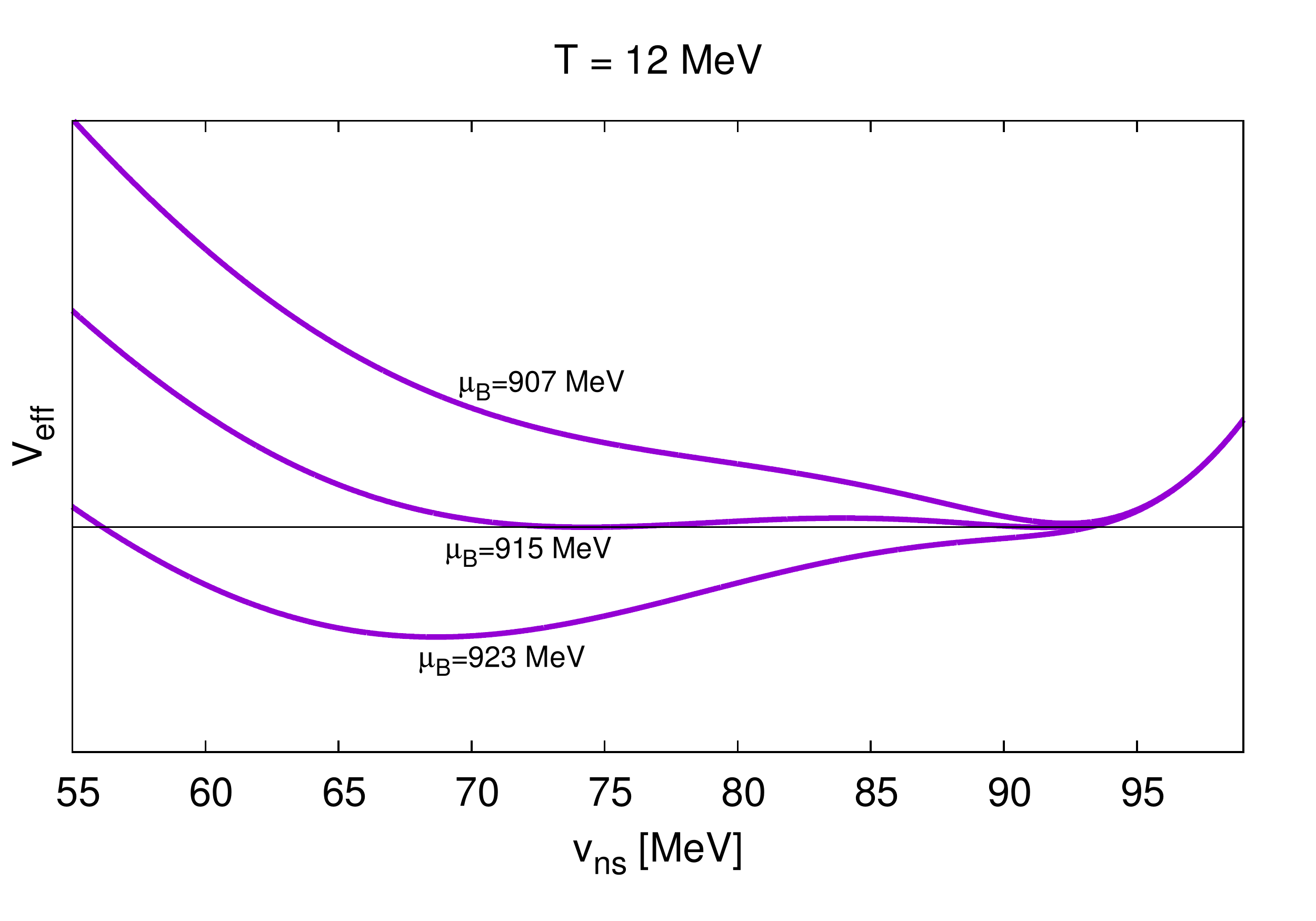}
\raisebox{-6.5cm}{
\includegraphics[bb = 0 550 520 170,scale=0.36,angle=270]{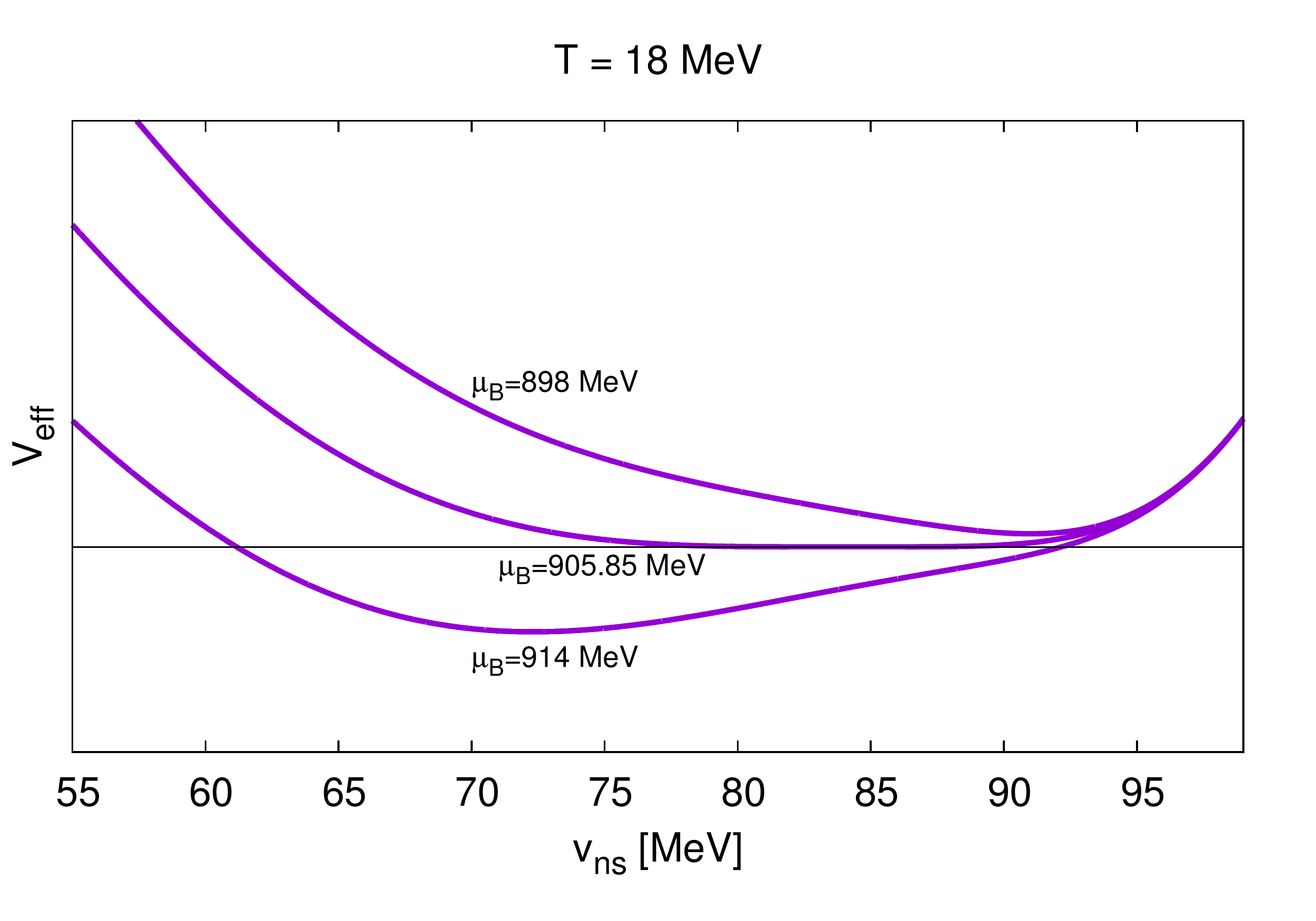}}
\caption{Shape of the effective potential for three different temperatures, as a function of $\mu_B$. The plots demonstrate how the critical end point is approached and how the first order transition is gradually turning into second order and a crossover.}
\label{fig1}
\end{center}
\end{figure*}

We still have to determine four more parameters (i.e. $m^2$, $g_1$, $g_2$, $a$) related to the three-flavor piece $V_k$ of $V_{\eff\!,k}$. The requirements here are only to reproduce physical masses \cite{pdg}. Using (\ref{Eq:Vefffinal}), we calculate all meson masses (see also Appendix B) and require the kaon, $\eta$, $\eta'$, and $a_0$ to get their physical values. Note that, as mentioned already, due to the construction of (\ref{Eq:Vefffinal}), the pion mass cannot be used to determine model parameters of the three-flavor piece, as due to cancellations of the first two lines, it is insensitive to the yet undetermined parameters. The following choices reproduce all the aforementioned masses within a $10\%$ accuracy compared to their physical value:
\bea
m^2&\approx &-0.95 \GeV^2, \quad g_1\approx 2.67 \nonumber\\
g_2&\approx &62.3, \quad a\approx -2.8 \GeV. 
\eea

It has to be noted that all the flow equations for $V_k(I_1)$, $C_k(I_1)$ and $A_k(I_1)$ are solved on a grid, with a step size of $\delta I_1=(10\MeV)^2$ in an interval of $I=[0:2]\GeV^2$. In $k$-space we initialize the flows at $\Lambda=1 \GeV$ and integrate down to $k=0$ with a step size $\delta k=10^{-2}\MeV$. Field derivatives are crucial to be calculated accurately, and we were using the seven-point formula (except for close to the boundaries, where five- and three-point formulas were employed). 

Also note that, one always needs to solve all equations at $T=0$, as even the finite temperature expression of $V_{\eff\!,k=0}$ contains the corresponding functions at $T=0$. Once this step is done, one recalculates the aforementioned functions at $T\neq 0$ to obtain the complete effective potential at any temperature. As for the masses, one needs to go through differentiations with respect to the field variables, which always come in through chiral invariants, as required by symmetry. They can be calculated with the help of some useful formulas that can be found in Appendix B. 

\begin{figure*}
\begin{center}
\raisebox{0.05cm}{
\includegraphics[bb = 0 490 520 770,scale=0.36,angle=270]{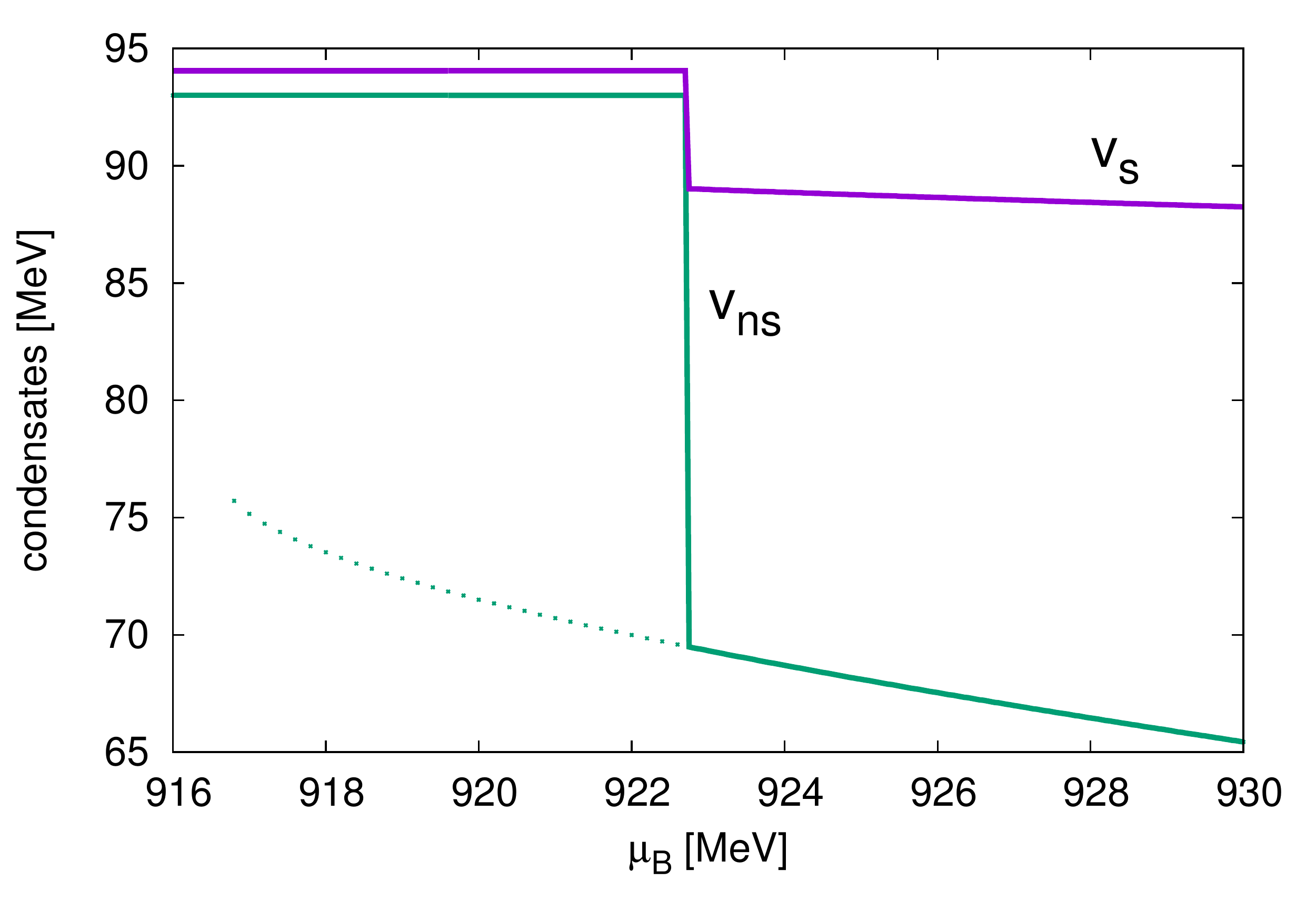}}
\includegraphics[bb = 0 80 520 170,scale=0.36,angle=270]{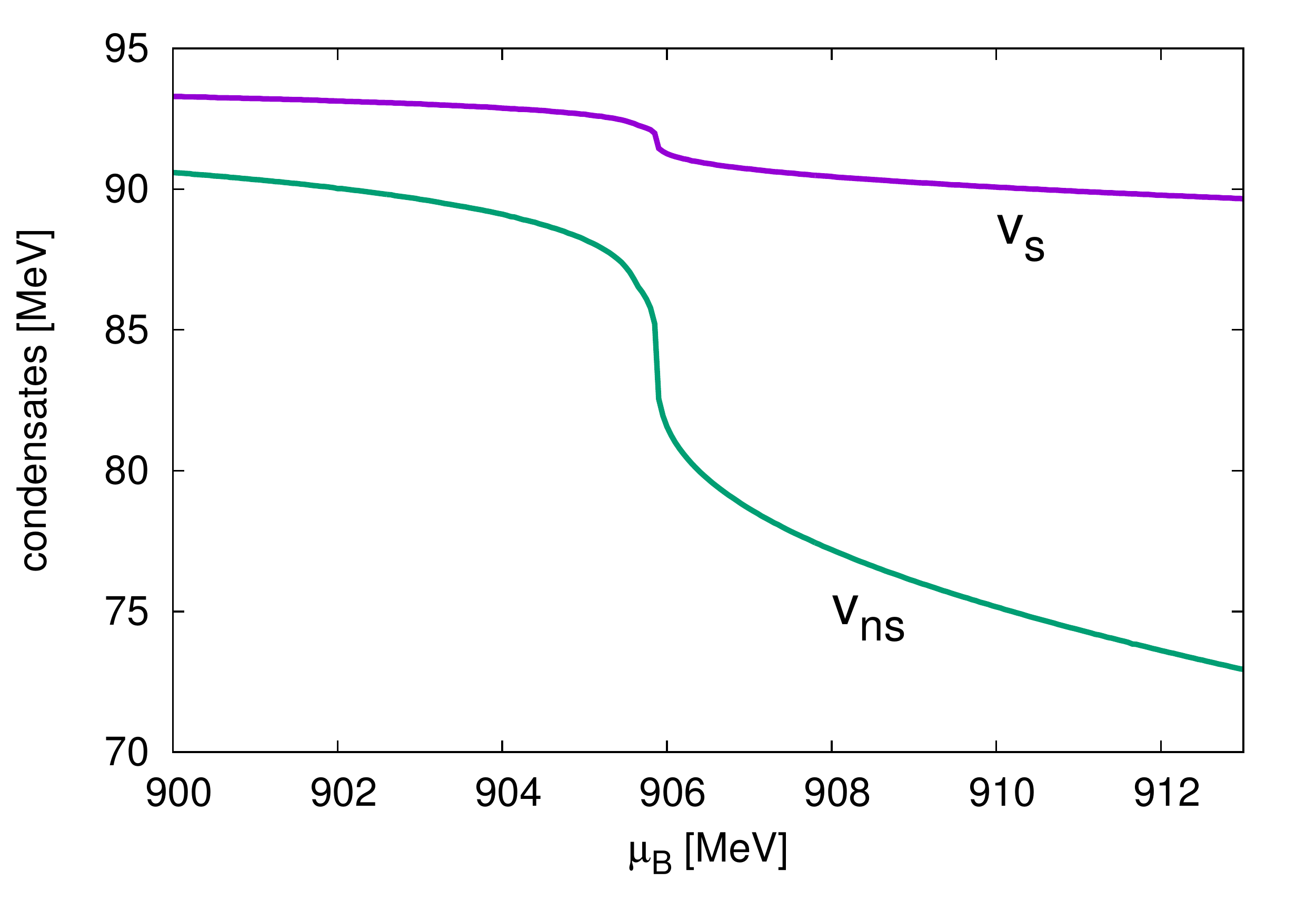}
\caption{Partial restoration of chiral symmetry due to the nuclear liquid-gas transition. We show how the condensates depend on $\mu_B$ at $T=0$ (left) and at $T=18$ MeV (right), which corresponds to the critical end point, i.e. a second order transition.}
\label{fig2}
\end{center}
\end{figure*}

\begin{figure*}
\begin{center}
\includegraphics[bb = 0 80 500 670,scale=0.44,angle=270]{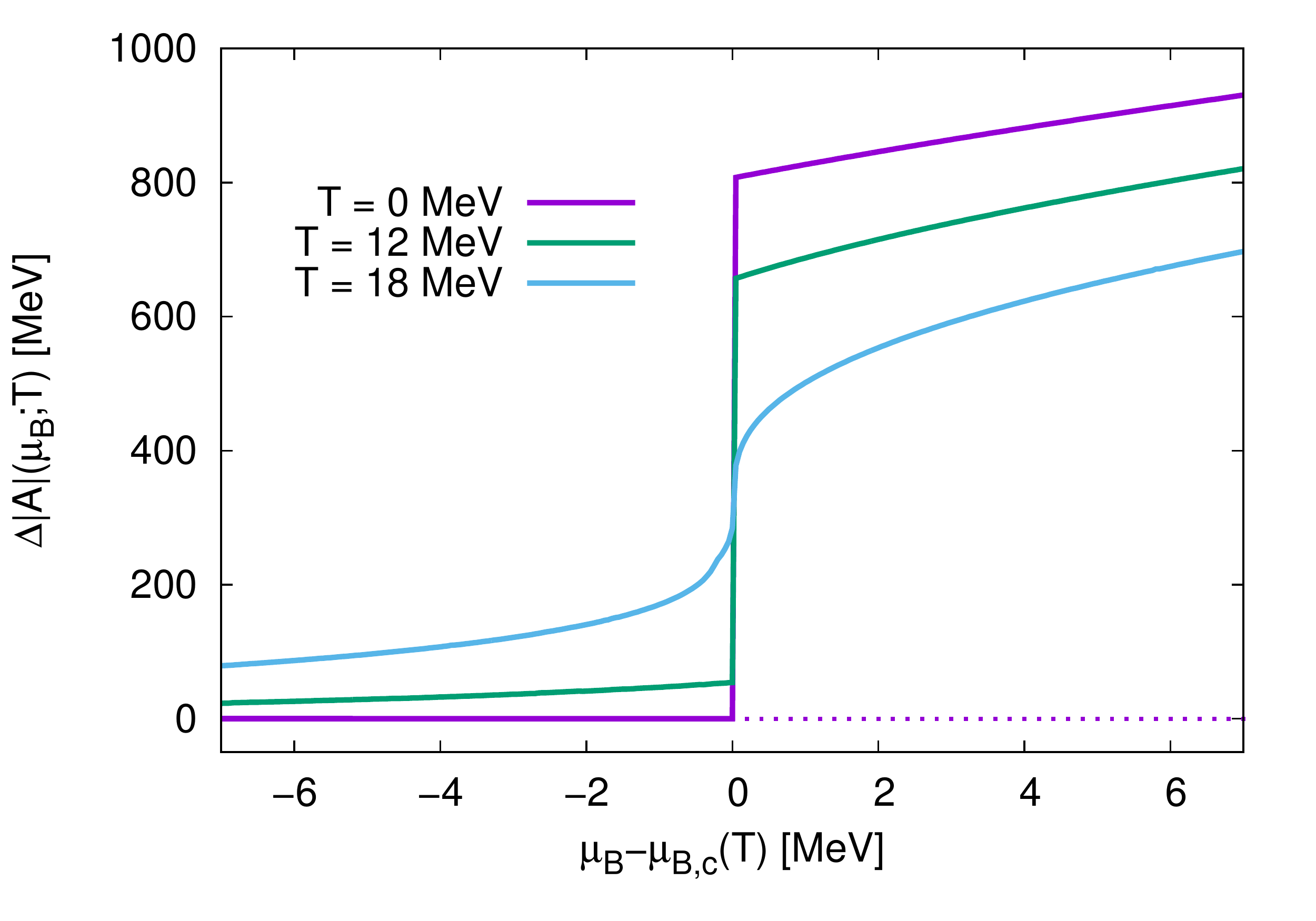}
\caption{Plot of the change in the strength of the anomaly in the minimum of the effective potential as a function of $\mu_B-\mu_{B,c}$, for different temperatures. Note that $\mu_{B,c}$ corresponds to the critical chemical potential and it depends on the temperature.}
\label{fig3}
\end{center}
\end{figure*}

\section{Results}

Now we review the results of the paper. In Fig. 1 the effective potential is shown around the liquid-gas transition as a function of the baryochemical potential for different temperatures. It is demonstrated how the first order transition is turning into second order and a crossover for $T>18 \MeV$. Related plots can also be found in Fig. 2, where the nonstrange and strange condensates are shown as a function of $\mu_B$ for $T=0$ and at $T=18 \MeV$. The latter belongs to the critical end point (i.e. a second order transition). It can be seen that even though nucleon fluctuations do not couple to the strange sector, the nonstrange condensate ``pulls'' the strange one toward a lower value as it changes.

In Fig. 3 we plot how the anomaly coefficient in the minimum of the effective potential, i.e. $A_{k=0}[I_1=(v_{\ns\!,\!\min}^2+v_{\s\!,\!\min}^2)/2]$ behaves at various temperatures as a function of $\mu_B-\mu_{B,c}$, where $\mu_{B,c}$ is the critical baryochemical potential at a given temperature. For practical reasons we define and plot an anomaly difference function,
\bea
\Delta |A|(\mu_B;T)=\Big|A_{k=0}|_{\mu_B}-A^{T=0}_{k=0}|_{\mu_B=0}\Big|.
\eea 
There is a clear tendency of strengthening, which can be understood by taking a look at Fig. 4 showing the $A_{k=0}(I_1)$ profile function \cite{fejos16} (note that it does not depend on $\mu_B$). First, we note that the change of $A_k(I_1)$ as a function is negligible in the temperature interval in question (i.e. $[0$:$18] \MeV$), but as $v_{\ns\!,\!\min}$ and $v_{\s\!,\!\min}$ decreases at the nuclear liquid-gas transition, the actual anomaly strength goes up. Notice how important the functional nature of our method is, as no condensate dependence of the anomaly function could have been obtained using conventional perturbation theory. The corresponding term, $A_k(I_1)\cdot I_{\det}$, can be interpreted as an infinite resummation of $I_1^n \cdot I_{\det}/n!$ operators. 

\begin{figure*}
\begin{center}
\includegraphics[bb = 0 80 500 670,scale=0.44,angle=270]{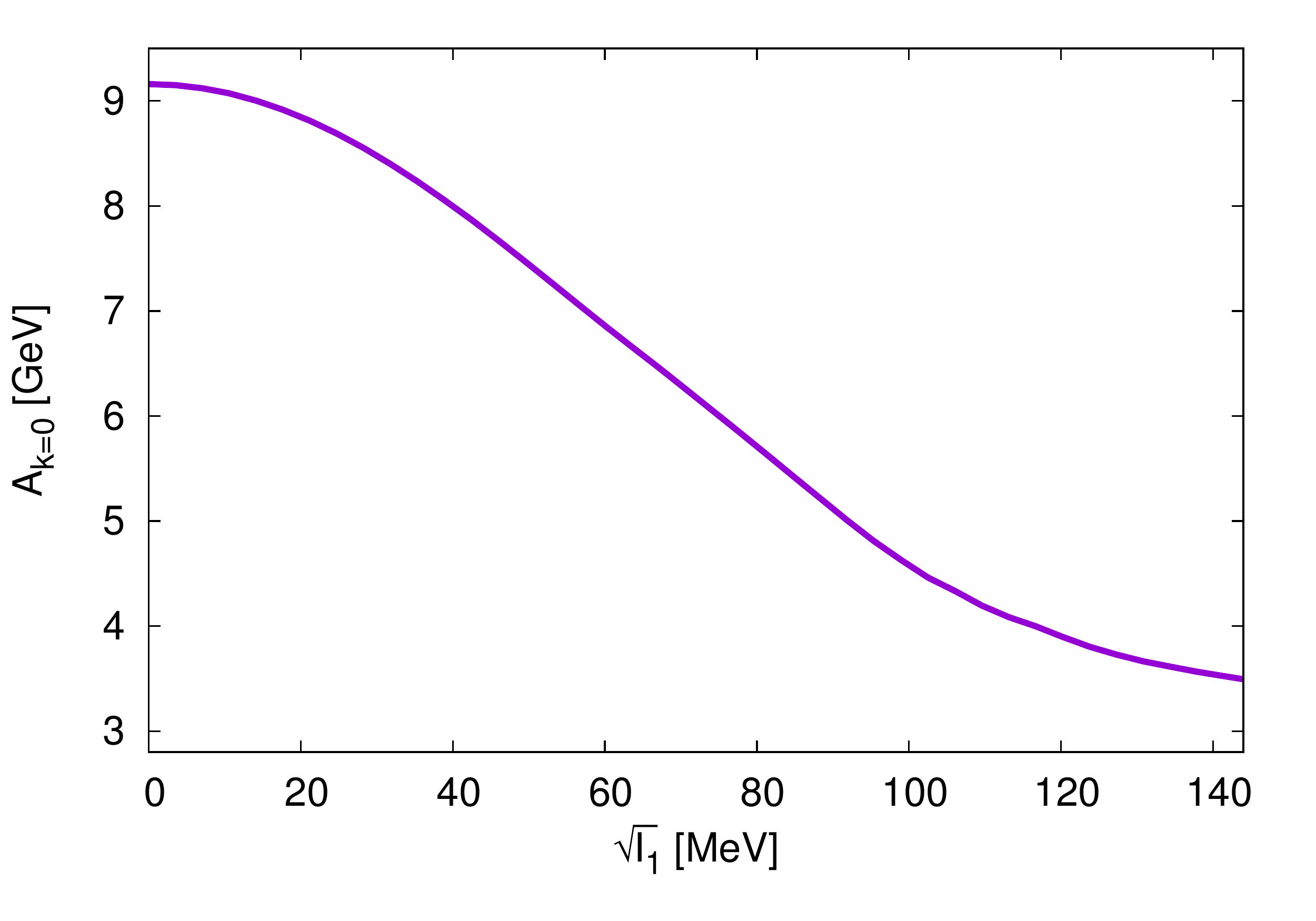}
\caption{Profile function $A_{k=0}$ at zero temperature, as a function of the chiral invariant $I_1|_{v_{\ns},v_{\s}}=(v_{\ns}^2+v_{\s}^2)/2$. In the temperature range that corresponds to a first order transition (i.e. $0\leq T \lesssim 18 \MeV$), change of the shape is not visible.}
\label{fig3}
\end{center}
\end{figure*}

\begin{figure*}
\begin{center}
\includegraphics[bb = 0 80 500 670,scale=0.44,angle=270]{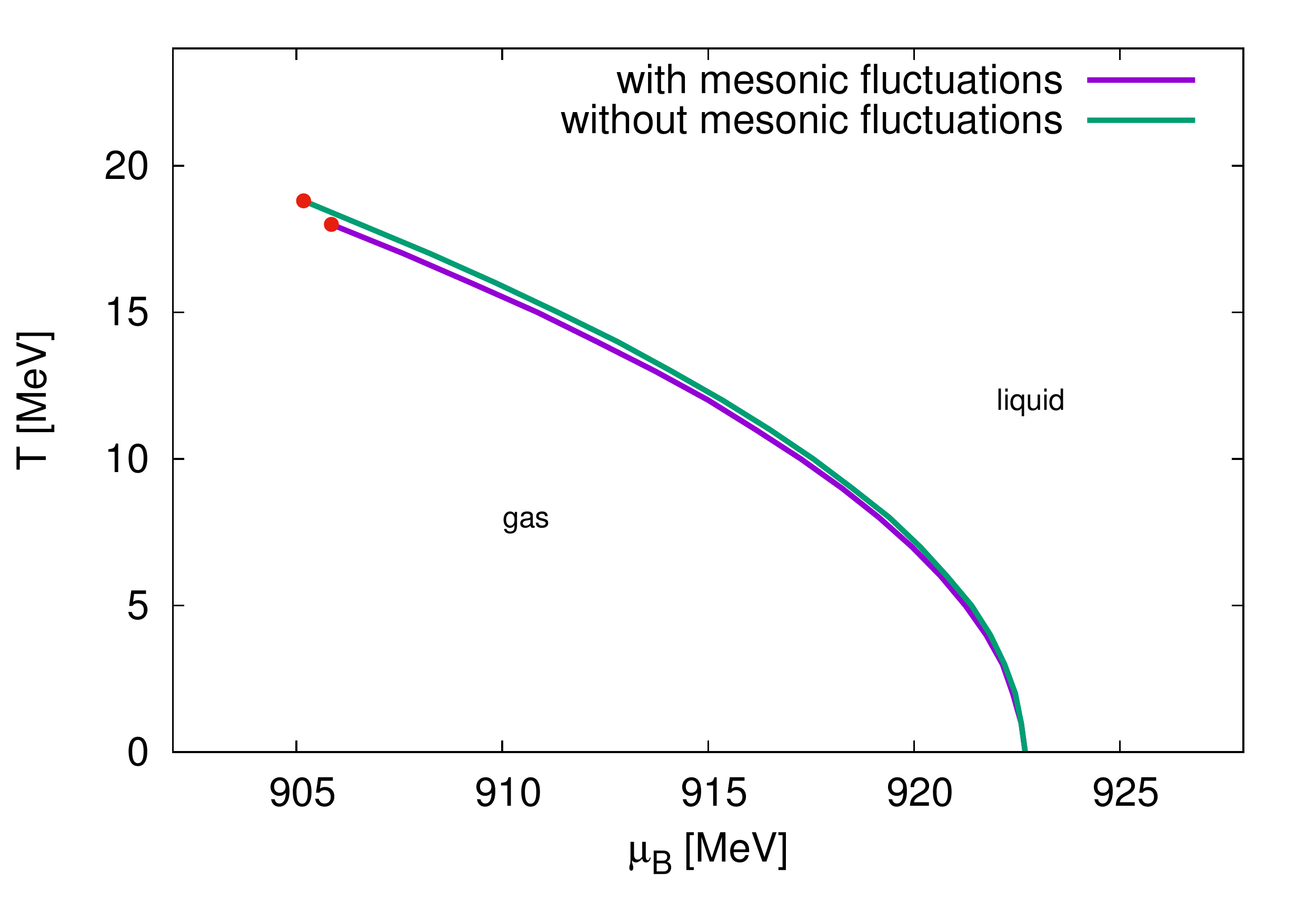}
\caption{Phase boundary in the $\mu_B-T$ plane. Once the end point is set via parametrization, the shape is not really sensitive to the inclusion of mesonic fluctuations.}
\label{fig5}
\end{center}
\end{figure*}

\begin{figure*}
\begin{center}
\includegraphics[bb = 0 80 500 670,scale=0.44,angle=270]{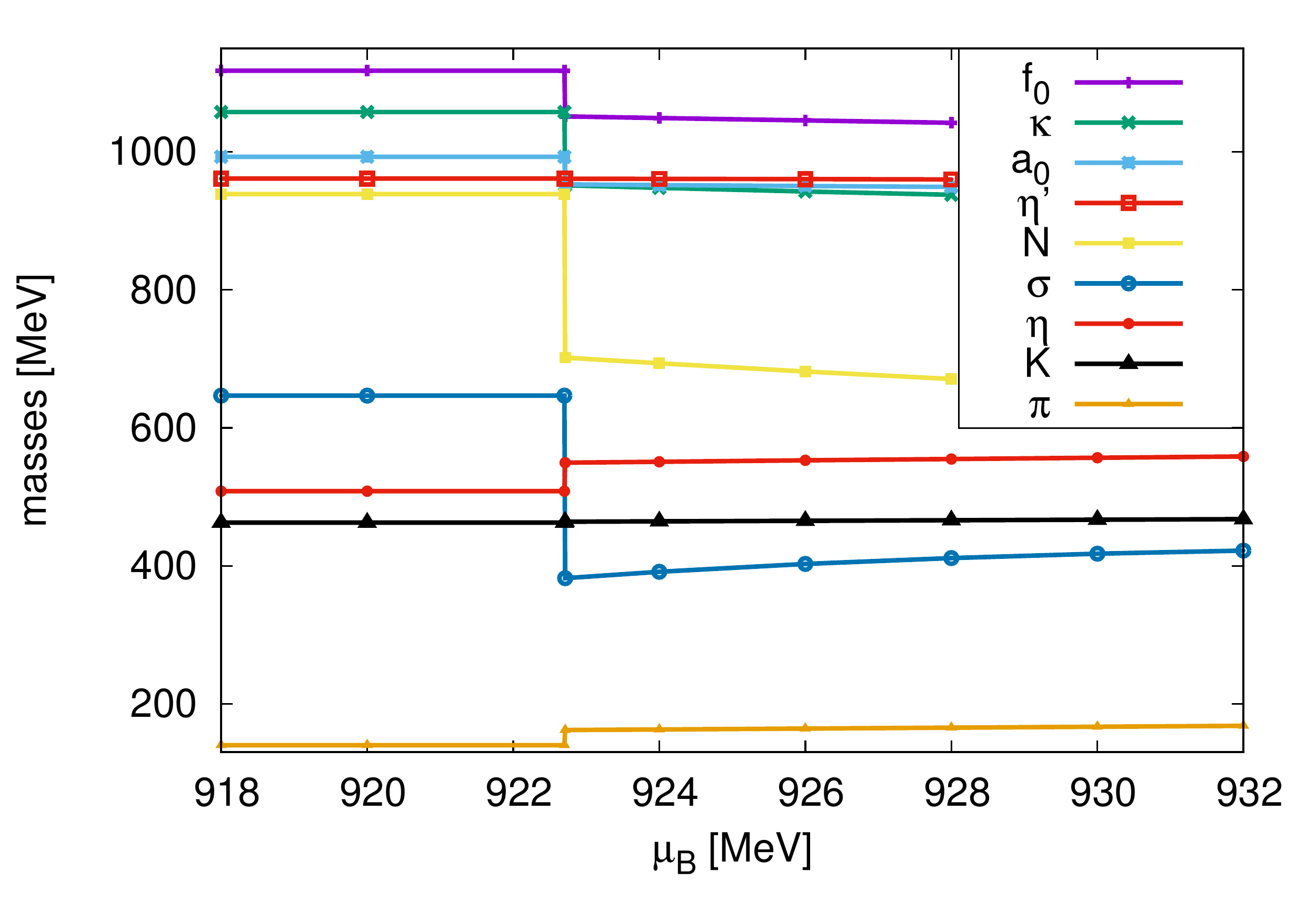}
\caption{Mass spectrum at zero temperature. Notice how the increasing anomaly flattens the $\eta'$ mass.}
\label{fig6}
\end{center}
\end{figure*}

\begin{figure*}
\begin{center}
\includegraphics[bb = 0 80 500 670,scale=0.44,angle=270]{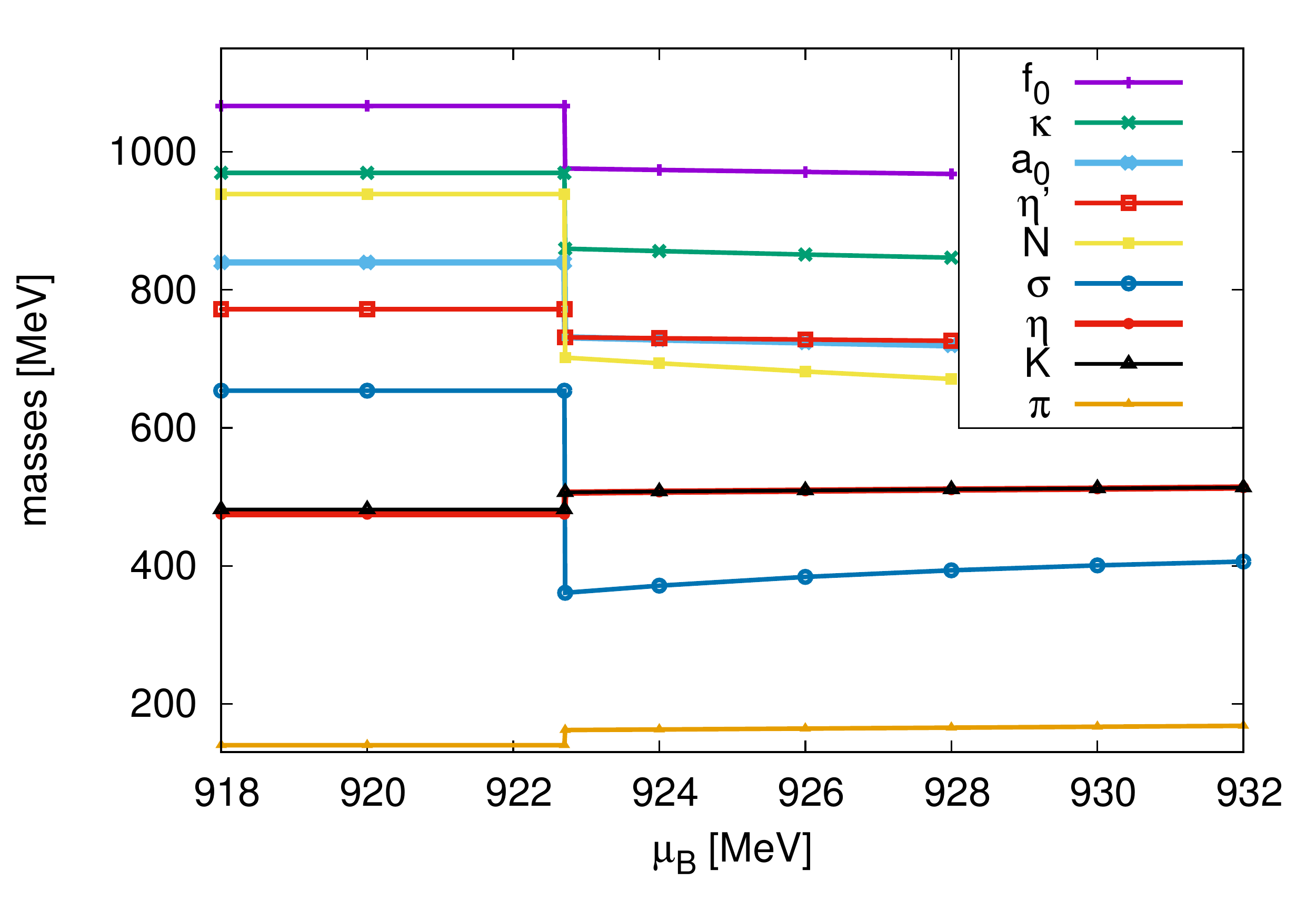}
\caption{Mass spectrum at zero temperature, without taking into account the field dependence of the anomaly coefficient (model parameters are the same as in Fig. 6).}
\label{fig7}
\end{center}
\end{figure*}

The next thing we are interested in is the phase boundary on the $\mu_B-T$ plane, which is shown in Fig. 5. As described in the previous subsection, through parametrization the end point is set to $18 \MeV$ \cite{elliott13}, and our calculations show that the curve is not really sensitive to the inclusion of mesonic fluctuations. This is appealing from the point of view that even a mean field calculation \cite{floerchinger12} (i.e. inclusion of fermionic one-loop effects only) is quite stable. Mesonic fluctuations seem to be only crucial from the point of view of the anomaly and the mass spectrum. The latter can be seen in Figs. 6 and 7. Figure 6 corresponds to the full calculation, where one notices a flat $\eta'$ mass, which is consistent with the earlier study \cite{fejos17}. This can be understood as follows. Even though the condensates abruptly decrease at the transition point, at the same time the anomaly goes up, therefore, ``anomaly''$\times$``condensate'' type of terms do not necessarily decrease: there is a competition between the increasing anomaly and the decreasing condensates. In our case it seems that the opposite effects almost exactly cancel each other leading to a flat $\eta'$ mass as a function of the chemical potential (or nuclear density).

For comparison, we plot in Fig. 7 a spectrum, where the anomaly coefficient is set to its vacuum value and not allowed to change as the condensates vary. In this case the $\eta'$ mass indeed decreases about ${\cal O}(10\%)$, as no change in the anomaly can compensate the drop of the chiral condensates. This result reproduces many earlier studies, which treated, as a somewhat crude approximation, the anomaly coefficient as a constant. Note that we have performed no reparametrization, thus masses deviate from their physical values. This shows that inhomogeneities of the anomaly function also carry significant contributions to the masses. 

\section{Summary}

In this paper we calculated the thermodynamic properties of the three-flavor linear sigma model extended with nucleon degrees of freedom. The nucleon-nucleon short-range interaction was modeled by a vector-particle $\omega$, and the method which we employed to calculate fluctuation effects was the functional renormalization group technique. We found that the coefficient of the determinant operators, which describe the $U_A(1)$ anomaly in the effective field theory setting, acquires field dependence due to mesonic fluctuations, and as a result, the melting of the nonstrange and strange condensates can cause elevation of the anomaly.

We saw that this indeed happens at finite temperature and density, in particular during the nuclear liquid-gas transition. Since the latter is of first order at low temperatures, a corresponding jump also takes place in the anomaly if $T\lesssim 18 \MeV$. As expected, this qualitatively changes the mass spectrum in the medium compared to earlier studies, e.g. we observed an $\eta'$ mass that did not drop at the phase transition, but stayed rather smooth as a function of the chemical potential (or nucleon density). If the spectrum in nuclear medium indeed shows such a behavior, an $\eta'$-nucleon bound state formation might be out of reach. We wish to point out that the obtained results are consistent with our earlier study that did not take into account any nucleon-nucleon interaction, and the corresponding liquid-gas transition. 

We also argued that mesonic fluctuations, while being crucial from the point of view of anomaly evolution and the mass spectrum, does not affect the critical end point of the liquid-gas transition. This shows that a mean-fieldlike approximation, where only one-loop fermionic contributions are considered \cite{floerchinger12,drews16}, is quite reliable, if one is interested in the thermodynamics of the nonstrange sector.

The study can be extended in various directions. First of all, we have not taken into account any instanton effect. The anomaly evolution is solely driven by mesonic fluctuations in this study, and one is interested in how the results would change if instantons were included via an environment dependent bare anomaly coefficient. Furthermore, isospin asymmetric nuclear matter could also be studied (it is of importance for neutron star physics), where the isovector $\rho$ particle also has to be introduced. Finally, a more complete study of the system would also include all the baryons, in particular hyperons, which is expected to be relevant at higher densities than we have studied in the present paper. This requires an extension of the model toward a complete flavor $SU(3)$ symmetry. These directions represent future works to be reported elsewhere.

\makeatletter
\@addtoreset{equation}{section}
\makeatother 

\renewcommand{\theequation}{A\arabic{equation}} 

\appendix
\section{Flow equations in the three-flavor sector}

The procedure of obtaining flow equations for $U_k(I_1)$, $C_k(I_1)$ and $A_k(I_1)$ is described in Sec. IIB. Here we list the corresponding results:
\begin{widetext}
\begin{subequations}
\label{Eq:3flflow}
\bea
\label{Eq:flow_Uk}
\partial_kU_k(U_1)&=&\frac{k^4T}{6\pi^2}\sum_{n=-\infty}^\infty\Bigg[\frac{9}{\Omega_n^2+E_\pi^2}+\frac{8}{\Omega_n^2+E_{a_0}^2}+\frac{1}{\Omega_n^2+E_\sigma^2}\Bigg], \\
\label{Eq:flow_Ck}
\partial_k C_k(I_1)&=&\frac{k^4T}{6\pi^2}\sum_{n=-\infty}^\infty\Bigg[\frac{4(3C_k+2I_1C_k')^2/3}{(\Omega_n^2+E_{a_0}^2)^2(\Omega_n^2+E_\sigma^2)}
+\frac{128C_k^5I_1^3/9}{(\Omega_n^2+E_\pi^2)^3(\Omega_n^2+E_{a_0}^2)^3}+\frac{24C_k\left(C_k-I_1C_k'\right)}{(\Omega_n^2+E_{a_0}^2)^3}\nonumber\\
&+&\frac{4\left(3C_kC_k'I_1+4I_1^2C_k'^2+C_k(3C_k-2C_k''I_1^2)\right)/3}{(\Omega_n^2+E_{a_0}^2)(\Omega_n^2+E_\sigma^2)^2}
+\frac{64C_k^3I_1^2(C_k-I_1C_k')/3}{(\Omega_n^2+E_\pi^2)^2(\Omega_n^2+E_{a_0}^2)^3}-\frac{48C_k^2I_1^2C_k'}{(\Omega_n^2+E_\pi^2)(\Omega_n^2+E_{a_0}^2)^3} \nonumber\\
&+&\frac{6C_k-17I_1C_k'}{(\Omega_n^2+E_{a_0}^2)^2}\frac{1}{I_1}-\frac{6C_k+9I_1C_k'+2I_1^2C_k''}{(\Omega_n^2+E_\sigma^2)^2}\frac{1}{I_1}
+\frac{4C_k(6C_k+9I_1C_k'+2I_1^2C_k'')/3}{(\Omega_n^2+E_{a_0}^2)(\Omega_n^2+E_\sigma^2)^2}\Bigg],\\
\label{Eq:flow_Ak}
\partial_k A_k(I_1)&=&\frac{k^4T}{6\pi^2}\sum_{n=-\infty}^\infty\Bigg[-\frac{9A_k'}{(\Omega_n^2+E_\pi^2)^2}-\frac{9A_k}{I_1(\Omega_n^2+E_\pi^2)^2}
-\frac{8A_k'}{(\Omega_n^2+E_{a_0}^2)^2}+\frac{12A_k}{I_1(\Omega_n^2+E_{a_0}^2)^2} \nonumber\\
&-&\frac{3A_k}{(\Omega_n^2+E_\sigma^2)^2I_1}+\frac{7A_k'}{(\Omega_n^2+E_\sigma^2)^2}+\frac{2I_1A_k''}{(\Omega_n^2+E_\sigma^2)^2}\Bigg],
\eea
\end{subequations}
\end{widetext}
where $\Omega_n=2\pi nT$, and
\begin{subequations}
\bea
E_\pi^2&=&k^2+U_k'(I_1), \\
E_{a_0}^2&=&k^2+U_k'(I_1)+\frac43 I_1 C_k(I_1), \\
E_{\sigma}^2&=&k^2+U_k'(I_1)+2I_1 U_k''(I_1).
\eea
\end{subequations}
Every Matsubara sum in Eq. (\ref{Eq:3flflow}) can be generated via taking derivatives of the functions
\bea
S^{(1)}(E)=T\sum_{n=-\infty}^{+\infty} \frac{1}{\Omega_n^2+E^2}=\frac{\coth(E/2T)}{2E},
\eea
and
\bea
S^{(2)}(E_1,E_2)&=&T\sum_{n=-\infty}^{+\infty} \frac{1}{(\Omega_n^2+E_1^2)(\Omega_n^2+E_2^2)}\nonumber\\
&=&\frac{1}{2E_1 E_2}\frac{E_1\coth(E_2/2T)-E_2\coth(E_1/2T)}{E_1^2-E_2^2}. \nonumber\\
\eea

\renewcommand{\theequation}{B\arabic{equation}} 
\section{Field derivatives}

Evaluating the flow equations and calculation of the mass spectrum require the determination of field derivatives of the effective potential. As described in Sec. IIC, the latter is a function of chiral invariants, and here we list various field derivatives of them.

For the sake of readability, we repeat some definitions:
\bea
I_1&=&\Tr(M^\dagger M), \quad I_2=\Tr\Big(M^\dagger M - \Tr(M^\dagger M)/3\Big)^2, \nonumber\\
I_{\det}&=&\det M^\dagger + \det M, \quad \tilde{I}_1=\Tr(\tilde{M}^\dagger \tilde{M}).
\eea
Note that [$T^a$ are $U(3)$, $\tilde{T}^a$ are $U(2)$ generators]
\bea
M=\sum_{a=0}^8 (s^a+i\pi^a)T^a
\eea
while
\bea
\tilde{M}=\sum_{a=\ns\!,1,2,3}(s^a+i\pi^a)\tilde{T}^a,
\eea
and in accordance, $I_1$, $I_2$ and $I_{\det}$ are invariant under the $U_L(3)\times U_R(3)$ group [$I_{\det}$ breaks $U_A(1)$ as it should], while $\tilde{I}_1$ shows $U_L(2)\times U_R(2)$ invariance only. 

We impose a background of
\bea
M|_{v_0,v_8}=v_0T^0+v_8T^8\equiv v_{\ns}T^{\ns}+v_{\s}T^{\s},
\eea
[for the transformation matrix between $(0,8)$ and (ns, s) see (\ref{Eq:snsss})] and correspondingly
\bea
\tilde{M}|_{v_0,v_8}=v_{\ns}\tilde{T}^{\ns}.
\eea
\begin{widetext}
In this background, the invariants are
\bea
I_1|_{v_0,v_8}&=&\frac{v_0^2+v_8^2}{2}, \hspace{7.45cm} I_2|_{v_0,v_8}=\frac{v_8^2}{24}(v_8-2\sqrt{2}v_0)^2, \\
I_{\det}\big|_{v_0,v_8}&=&\frac{1}{3\sqrt6}(v_0^3-\frac32v_0v_8^2-\frac{1}{\sqrt2}v_8^3), \hspace{4.6cm} \tilde{I}_1\big|_{v_0,v_8}=\frac13\Big(v_0+\frac{1}{\sqrt{2}}v_8\Big)^2,
\eea
while their derivatives turn out to be
\bea
\frac{\partial I_1}{\partial s^a}\bigg|_{v_0,v_8}&=&v_0\delta^{a0}+v_8\delta^{a8}, \hspace{6.25cm} \frac{\partial I_1}{\partial \pi^a}\bigg|_{v_0,v_8}=0, \\
\frac{\partial I_2}{\partial s^a}\bigg|_{v_0,v_8}&=&\left(\frac{2v_0v_8^2}{3}-\frac{1}{3\sqrt{2}}v_8^3\right)\delta^{a0}+\left(\frac{2v_0^2v_8}{3}-\frac{v_0v_8^2}{\sqrt{2}}+\frac{v_8^3}{6}\right)\delta^{a8}, \hspace{0.41cm} \frac{\partial I_2}{\partial \pi^a}\bigg|_{v_0,v_8}=0, \\
\frac{\partial I_{\det}}{\partial s^a}\bigg|_{v_0,v_8}&=&\frac{2v_0^2-v_8^2}{2\sqrt{6}}\delta^{a0}-\frac{v_8(\sqrt{2}v_0+v_8)}{2\sqrt{3}}\delta^{a8}, \hspace{3.2cm} \frac{\partial I_{\det}}{\partial \pi^a}\bigg|_{v_0,v_8}=0, \\
\frac{\partial \tilde{I}_1}{\partial s^a}\bigg|_{v_0,v_8}&=&\frac{\sqrt2}{3}(\sqrt2 v_0+v_8)\delta^{a0}+\frac13(\sqrt2 v_0+v_8)\delta^{a8} \hspace{2.7cm} \frac{\partial \tilde{I}_1}{\partial \pi^a}\bigg|_{v_0,v_8}=0.
\eea
The second derivatives are
\bea
\frac{\partial^2 I_1}{\partial s^a \partial s^b}\bigg|_{v_0,v_8}&=&\delta^{ab}, \hspace{7.3cm} \frac{\partial^2 I_1}{\partial \pi^a \partial \pi^b}\bigg|_{v_0,v_8}=\delta^{ab}, \\
\frac{\partial^2 I_2}{\partial s^a s^b}\bigg|_{v_0,v_8}&=&
\begin{cases}
\frac{2}{3}v_8^2,  \hspace{4.5cm} \ife \hspace{0.1cm} a=b=0\\
-\frac{v_8^2}{\sqrt{2}}+\frac{4}{3}v_0v_8,  \hspace{3.0cm} \ife \hspace{0.1cm} a=0,\hspace{0.1cm} b=8 \hspace{0.1cm} \orr \hspace{0.1cm} a=8,\hspace{0.1cm} b=0\\
\frac{2}{3}v_0^2+\frac{v_8^2}{2}-\sqrt{2}v_0v_8,  \hspace{2.1cm} \ife \hspace{0.1cm} a=b=8\\
\frac{2}{3}v_0^2+\frac{v_8^2}{6}+\sqrt{2}v_0v_8, \hspace{2.1cm} \ife \hspace{0.1cm} a=b=1,2,3\\
\frac{2}{3}v_0^2+\frac{v_8^2}{6}-\frac{1}{\sqrt{2}}v_0v_8, \hspace{2.15cm} \ife \hspace{0.1cm} a=b=4,5,6,7\\
0, \hspace{4.9cm}  \els
\end{cases}
\eea
\bea
\frac{\partial^2 I_2}{\partial \pi^a \pi^b}\bigg|_{v_0,v_8}&=&
\begin{cases}
0, \hspace{4.9cm} \ife \hspace{0.1cm} a=b=0\\
-\frac{v_8^2}{3\sqrt{2}}+\frac{2}{3}v_0v_8, \hspace{2.9cm} \ife \hspace{0.1cm} a=0,\hspace{0.1cm} b=8 \hspace{0.1cm} \orr \hspace{0.1cm} a=8,\hspace{0.1cm} b=0\\
\frac{v_8^2}{6}-\frac{\sqrt{2}}{3}v_0v_8, \hspace{3.18cm} \ife \hspace{0.1cm} a=b=8\\
-\frac{v_8^2}{6}+\frac{\sqrt{2}}{3}v_0v_8, \hspace{2.88cm} \ife \hspace{0.1cm} a=b=1,2,3\\
\frac{5}{6}v_8^2-\frac{1}{3\sqrt{2}}v_0v_8, \hspace{2.83cm} \ife \hspace{0.1cm} a=b=4,5,6,7\\
0, \hspace{4.9cm}  \els
\end{cases}
\eea
\bea
\frac{\partial^2 I_{\det}}{\partial s_i s_j}\bigg|_{v_0,v_8}&=&
\begin{cases}
\sqrt{\frac23}v_0,  \hspace{4.6cm} \ife \hspace{0.1cm} i=j=0\\
-\frac{v_8}{\sqrt6},  \hspace{4.8cm} \ife \hspace{0.1cm} i=0,\hspace{0.1cm} j=8 \hspace{0.1cm} \orr \hspace{0.1cm} i=8,\hspace{0.1cm} j=0\\
-\frac{v_0}{\sqrt6}-\frac{v_8}{\sqrt3},  \hspace{3.9cm} \ife \hspace{0.1cm} i=j=8\\
-\frac{v_0}{\sqrt6}+\frac{v_8}{\sqrt3}, \hspace{3.9cm} \ife \hspace{0.1cm} i=j=1,2,3\\
-\frac{v_0}{\sqrt6}-\frac{v_8}{2\sqrt3}, \hspace{3.75cm} \ife \hspace{0.1cm} i=j=4,5,6,7\\
0, \hspace{5.3cm}  \els
\end{cases}\\
\frac{\partial^2 I_{\det}}{\partial \pi^a \pi^b}\bigg|_{v_0,v_8}&=&
\begin{cases}
-\sqrt{\frac23}v_0,  \hspace{4.35cm} \ife \hspace{0.1cm} a=b=0\\
\frac{v_8}{\sqrt6},  \hspace{5.05cm} \ife \hspace{0.1cm} a=0,\hspace{0.1cm} b=8 \hspace{0.1cm} \orr \hspace{0.1cm} a=8,\hspace{0.1cm} b=0\\
\frac{v_0}{\sqrt6}+\frac{v_8}{\sqrt3},  \hspace{4.15cm} \ife \hspace{0.1cm} a=b=8\\
\frac{v_0}{\sqrt6}-\frac{v_8}{\sqrt3}, \hspace{4.15cm} \ife \hspace{0.1cm} a=b=1,2,3\\
\frac{v_0}{\sqrt6}+\frac{v_8}{2\sqrt3}, \hspace{4.0cm} \ife \hspace{0.1cm} i=j=4,5,6,7\\
0, \hspace{5.3cm}  \els
\end{cases}\\
\frac{\partial^2 \tilde{I}_{1}}{\partial s^a s^b}\bigg|_{v_0,v_8}&\equiv&\frac{\partial^2 \tilde{I}_{1}}{\partial \pi^a \pi^b}\bigg|_{v_0,v_8}=
\begin{cases}
\frac23,  \hspace{3.0cm} \ife \hspace{0.1cm} a=b=0\\
\frac{\sqrt{2}}{3},  \hspace{2.75cm} \ife \hspace{0.1cm} a=0,\hspace{0.1cm} b=8 \hspace{0.1cm} \orr \hspace{0.1cm} a=8,\hspace{0.1cm} b=0\\
\frac13,  \hspace{3.0cm} \ife \hspace{0.1cm} a=b=8\\
1, \hspace{3.05cm} \ife \hspace{0.1cm} a=b=1,2,3\\
0. \hspace{3.1cm}  \els
\end{cases}
\eea
\end{widetext}


\begin{thebibliography}{9}

\bibitem{borsanyi15} S. Borsanyi {\it et al.}, Phys. Lett. B{\bf 751}, 559 (2015).
\bibitem{demorest10} P. B. Demorest {\it et al.}, Nature (London) {\bf 467}, 1081 (2010).
\bibitem{antoniadis13} J. Antoniadis {\it et al.}, Science {\bf 340}, 6131 (2013).
\bibitem{abbott17} B. P. Abbott {\it et al.}, Phys. Rev. Lett. {\bf 119}, 161101 (2017).
\bibitem{tanaka18} Y. K. Tanaka {\it et al.}, Phys. Rev. C{\bf 97}, 015202 (2018).
\bibitem{costa05} P. Costa, M. C. Ruivo, C. A. de Sousa, and Yu. L. Kalinovsky, Phys. Rev. D{\bf 71}, 116002 (2005).
\bibitem{nagahiro06} H. Nagahiro, M. Takizawa, and S. Hirenzaki, Phys. Rev.
C{\bf 74}, 045203 (2006).
\bibitem{sakai13} S. Sakai and D. Jido, Phys. Rev. C{\bf 88}, 064906 (2013).
\bibitem{sakai16} S. Sakai and D. Jido, Prog. Theor. Exp. Phys. 013D01 (2017).
\bibitem{nuramatsu13} N. Muramatsu {\it et al.}, arXiv:1307.6411.
\bibitem{j-parc-hi16} J-PARC Heavy-Ion Program White Paper, http://silver.j-parc.jp/sako/white-paper-v1.21.pdf
\bibitem{parganlija13} D. Parganlija, P. Kovacs, Gy. Wolf, F. Giacosa, and D.-H. Rischke, Phys. Rev. D{\bf 87}, 014011 (2013).
\bibitem{herbst14} T. K. Herbst, M. Mitter, J.-M. Pawlowski, B.-J. Schaefer, and R. Steile, Phys. Lett. B{\bf 731}, 248 (2014).
\bibitem{mitter14} M. Mitter and B.-J. Schaefer, Phys. Rev. D{\bf 89}, 054027 (2014).
\bibitem{rennecke16} F. Rennecke and B.-J. Schaefer, Phys. Rev. D{\bf 96}, 016009 (2017).
\bibitem{kovacs16} P. Kovacs, Zs. Szep, and Gy. Wolf, Phys. Rev. D{\bf 93}, 114014 (2016).
\bibitem{almasi17} G. A. Almasi, B. Friman, and K. Redlich, Phys. Rev. D{\bf 96}, 014027 (2017).
\bibitem{aoki18} K.-I. Aoki, S.-I. Kumamoto, and M. Yamada, Nucl. Phys. B{\bf 931}, 105 (2018).
\bibitem{floerchinger12} S. Floerchinger and C. Wetterich, Nucl. Phys. A{\bf 890}$-${\bf 891}, 11 (2012). 
\bibitem{drews16} M. Drews and W. Weise, Prog. Part. Nucl. Phys. {\bf 93}, 69 (2017).
\bibitem{eichmann16} G. Eichmann, C.-S. Fischer, and C.-A. Welzbacher, Phys. Rev. D{\bf 93} 034013 (2016).
\bibitem{marko13} G. Marko, U. Reinosa, and Z. Szep, Phys. Rev. D{\bf 87}, 105001 (2013).
\bibitem{marko15} G. Marko, U. Reinosa, and Z. Szep, Phys. Rev. D{\bf 92}, 125035 (2015).
\bibitem{berges02} J. Berges, N. Tetradis, and C. Wetterich, Phys. Rep. {\bf 363}, 223 (2002). 
\bibitem{kopietz} P. Kopietz, L. Bartosch and F. Sch\"utz, Introduction to the Functional Renormalization Group, Lect. Notes Phys. 798, Springer, 2010.
\bibitem{fejos16} G. Fejos and A. Hosaka, Phys. Rev. D{\bf 94}, 036005 (2016).
\bibitem{fejos17} G. Fejos and A. Hosaka, Phys. Rev. D{\bf 95}, 116011 (2017). 
\bibitem{schaefer98} T. Schaefer and E. Shuryak, Rev. Mod. Phys. {\bf 70}, 323 (1998).
\bibitem{fejos15} G. Fejos, Phys. Rev. D{\bf 92}, 036011 (2015).
\bibitem{berges97} J. Berges and C. Wetterich, Nucl. Phys. B{\bf 487}, 675 (1997).
\bibitem{fukushima10} K. Fukushima, K. Kamikado, and B. Klein, Phys. Rev. D{\bf 83}, 116005 (2011).
\bibitem{pdg} Particle Data Group, http://pdglive.lbl.gov/Viewer.action.
\bibitem{stone14} J. R. Stone, N. J. Stone, and S. A. Moszkowski, Phys. Rev. C{\bf 89}, 044316 (2014).
\bibitem{elliott13} J. B. Elliott, P. T. Lake, L. G. Moretto, and L. Phair, Phys. Rev. C{\bf 87}, 054622 (2013).
\end{thebibliography}
\end{document}